\documentclass[aps,prx,twocolumn,superscriptaddress,raggedbottom,showpacs,floatfix,longbibliography,nofootinbib]{revtex4-2}
\usepackage{bm}
\usepackage{physics}
\usepackage{amssymb}
\usepackage{amsmath}
\usepackage[usenames,dvipsnames]{xcolor}
\usepackage{soul}
\usepackage{multirow}
\usepackage{mathtools}
\usepackage{tikz}
\usepackage[colorlinks, linkcolor = blue, citecolor = blue, urlcolor=blue, pdfborder={0 0 0 [0 0]},bookmarks=false]{hyperref}

\usepackage{lipsum}
\usepackage{manfnt}
\usepackage{enumerate}
\usepackage[export]{adjustbox}
\usepackage{wasysym}
\usepackage{algorithm}
\usepackage{algpseudocode}

\newcommand{\rc}[1]{{\color{red}{#1}}}
\newcommand{\bc}[1]{{\color{blue}{#1}}}
\newcommand{\gc}[1]{{\color{ForestGreen}{#1}}}

\usepackage{soul}

\def\Rr{\rc{$r$}}
\def\Bb{\bc{$b$}}
\def\Gg{\gc{$g$}}

\def\la{\left \langle}
\def\ra{\right \rangle}
\def\lp{\left (}
\def\rp{\right )}

\newcommand{\rXX}{{\color{red}{$rXX$}}}

\def\gXX{\gc{$gXX$}}
\def\bXX{\bc{$bXX$}}
\def\rZZ{\rc{$rZZ$}}

\def\gZZ{\gc{$gZZ$}}

\def\bZZ{\bc{$bZZ$}}
\def\rXXX{\rc{$rXXX$}}
\def\bXXX{\bc{$bXXX$}}
\def\rZZZ{\rc{$rZZZ$}}
\def\bZZZ{\bc{$bZZZ$}}

\def\Xeff{\check{X}}
\def\Zeff{\check{Z}}

\def\cuberX{\mancube$_\rc{r}(X)$}

\def\sqgX{{$\diamond_\gc{g}(X)$}}
\def\cubebX{\mancube$_\bc{b}(X)$}

\def\cuberZ{\mancube$_\rc{r}(Z)$}
\def\sqgZ{$\diamond_\gc{g}(Z)$}
\def\cubebZ{\mancube$_\bc{b}(Z)$}

\def\PrX{$P_{\rc{r}}(X)$}
\def\PgX{$P_{\gc{g}}(X)$}
\def\PbX{$P_{\bc{b}}(X)$}

\def\PrZ{$P_{\rc{r}}(Z)$}
\def\PgZ{$P_{\gc{g}}(Z)$}
\def\PbZ{$P_{\bc{b}}(Z)$}

\def\Pr{$P_{\rc{r}}$}
\def\Pg{$P_{\gc{g}}$}
\def\Pb{$P_{\bc{b}}$}

\makeatletter
\def\l@subsection#1#2{}
\def\l@subsubsection#1#2{}
\makeatother

\newcommand{\be}{\begin{equation}}
\newcommand{\ee}{\end{equation}}

\newcommand{\octahedron}{
   \raisebox{-4pt}{\includegraphics[height=3ex]{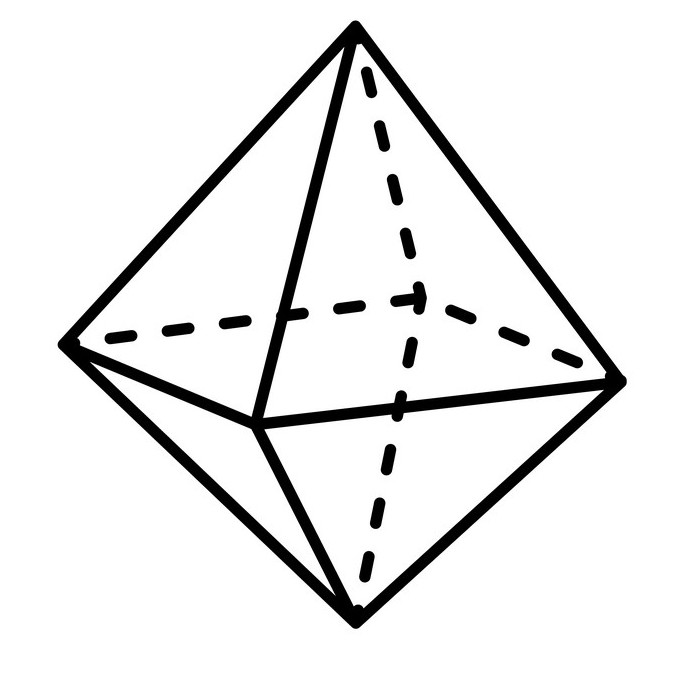}}}
   
\begin{document}

\title{Floquet codes without parent subsystem codes}
\author{Margarita Davydova}
\affiliation{Department of Physics, Massachusetts Institute of Technology, Cambridge, MA 02139, USA}
\affiliation{Kavli Institute for Theoretical Physics, University of California, Santa Barbara, California 93106, USA}
\author{Nathanan Tantivasadakarn}
\affiliation{Walter Burke Institute for Theoretical Physics and Department of Physics, California Institute of Technology, Pasadena, CA 91125, USA}
\affiliation{Department of Physics, Harvard University, Cambridge, MA 02138, USA}
\author{Shankar Balasubramanian}
\affiliation{Center for Theoretical Physics, Massachusetts Institute of Technology, Cambridge, MA 02139, USA}

\begin{abstract}
We propose a new class of error-correcting dynamic codes in two and three dimensions that has no explicit connection to any parent subsystem code. The two-dimensional code, which we call the CSS honeycomb code, is geometrically similar to that of the honeycomb code by Hastings and Haah, and also dynamically embeds an instantaneous toric code. However, unlike the honeycomb code it possesses an explicit CSS structure and its gauge checks do not form a subsystem code. Nevertheless, we show that our dynamic protocol conserves logical information and possesses a threshold for error correction. We generalize this construction to three dimensions and obtain a code that fault-tolerantly alternates between realizing two type-I fracton models, the checkerboard and the X-cube model. Finally, we show the compatibility of our CSS honeycomb code protocol and the honeycomb code by showing the possibility of randomly switching between the two protocols without information loss while still measuring error syndromes. We call this more general aperiodic structure  `dynamic tree codes', which we also generalize to three dimensions.  We construct a probabilistic finite automaton prescription that generates dynamic tree codes correcting any single-qubit Pauli errors and can be viewed as a step towards the development of practical fault-tolerant random codes.
\end{abstract}

\maketitle
\tableofcontents

\section{Introduction}

Any route towards new fault-tolerant schemes for quantum computing involves finding qualitatively different ways of performing quantum error correction.  A recent approach called operator quantum error correction~\cite{PhysRevLett.95.230504,PhysRevA.73.012340,PhysRevA.81.032301}requires one to recover only a part of the original `logical' state, while errors are allowed to affect the rest of it, which is spanned by `gauge qubits'. This can be accomplished by constructing  a \emph{subsystem code}, which is specified by a gauge group $\mathcal G$ that is generically a non-Abelian subgroup of the Pauli group. The stabilizer group $\mathcal S$ of the subsystem code is given by the centralizer of the gauge group, i.e. the set of the elements in the gauge group that commute with all elements of the group, and the logical qubits of the stabilizer code are split into the logical qubits of the subsystem code and gauge qubits, which are no longer used for encoding logical information.  Subsystem codes thus provide a generalization of the concept of stabilizer codes~\cite{Gottesman_1998}. 

In subsystem codes, syndrome measurement can be performed using generators of the gauge group only, which are usually low-weight (non-commuting) operators. This makes subsystem codes attractive for achieving fault tolerance and gives rise to several new proposals for realization of universal quantum computing.  A central idea in these proposals is a procedure called \emph{gauge fixing}, which corresponds to measuring a commuting subset of gauge operators (``checks''), thus fixing the states of the gauge qubits. The measured gauge operators are then added to the stabilizer  $\mathcal S$ of the subsystem code defined by the gauge group $\mathcal G$. Different ways of performing gauge fixing allows one to switch between different stabilizer codes $\mathcal S_1$ and $\mathcal S_2$ starting from the same parent gauge group $\mathcal G$.  This is aptly called `code switching' and a universal transversal set of gates can been realized this way from the gauge color code~\cite{PhysRevA.91.032330,bombin2015gauge}, quantum Reed-Muller code \cite{PhysRevLett.113.080501}, and more~\cite{PhysRevLett.111.090505}.   Furthermore, other methods that allow one to overcome the Eastin-Knill no-go theorem \cite{PhysRevLett.102.110502,PhysRevResearch.4.013092}, such as lattice surgery and code deformation \cite{bombin2009quantum,horsman2012surface}, can be unified into the framework of gauge fixing~\cite{vuillot2019code}.

Recently, a new dynamic error-correcting code, comprised of a time-periodic sequence of two-qubit Pauli measurements, was proposed by Hastings and Haah~\cite{hastings2021dynamically,Haah_2022}and dubbed the `honeycomb code'. It is considered the first example of a \emph{Floquet code}because of the inherent time periodicity in the measurement protocol. The honeycomb code is based on a subsystem code with a gauge group generated by terms in the Hamiltonian of the Kitaev honeycomb model \cite{kitaev2006anyons}. Notably, this subsystem code stabilizes no logical qubits  \cite{suchara2011constructions}.  However, the honeycomb code remedies this and dynamically generates logical qubits by measuring a commuting subset of the gauge group at each round, which constitutes one-third of the full set of two-qubit Pauli checks. This dynamic protocol generates a different stabilizer group at each instant in time which differ from that of the original subsystem code. In particular, the instantaneous stabilizer group of the dynamic code is equivalent to that of a toric code \cite{kitaev2003fault}on a certain superlattice, and the embedded code changes with period 3 while conserving logical information. Remarkably, the honeycomb code was also shown to possess a threshold \cite{Gidney2021faulttolerant,gidney2022benchmarking}.   From the quantum matter perspective, the honeycomb code not only switches between different realizations of $\mathbb Z_2$ topological order but also exhibits a kind of time crystalline behavior -- while the period of the cycling is 3, the period of the code is 6 because after 3 rounds an $e$/$m$ automorphism occurs.  This idea has been more generally explored in ref.~[\onlinecite{aasen}].

In this paper, we propose a new class of Floquet codes in two and three dimensions that are not based on parent subsystem codes. Our 2D construction is geometrically similar to that of the honeycomb code, but possesses an explicit CSS structure; therefore we call our code the \emph{CSS honeycomb code}. We show that this code embeds an instantaneous toric code, conserves logical information and possesses a threshold for error correction. It also turns out that the CSS honeycomb code performs an automorphism every three rounds.   Our 3D construction embeds two distinct type-I fracton models: we show that it cycles between realizing instances of checkerboard and X-cube  models \cite{PhysRevB.94.235157}while preserving logical information and being error-correcting as well. This is the first Floquet code we are aware of that prepares and cycles between fracton stabilizer codes. 

We argue that our 2D code cannot be reduced to the honeycomb code. However, we show that it is possible to fault-tolerantly switch between our CSS protocol and the honeycomb protocol. Moreover, we consider random disturbances of the protocol in time, thus generalizing Floquet codes to a large class of monitored random circuit codes which we call \emph{dynamic tree codes}, as the path of a single instance of such a code is a branch of the history tree of a probabilistic process. We show that a special class of these codes, i.e. \emph{random-flavor Floquet codes}, is fault-tolerant. Next, we construct a probabilistic finite automaton (PFA) that allows one to generate instances of dynamic tree codes that allow detection and correction of any single-qubit Pauli error.  We conjecture that a large class of PFA-generated dynamic tree codes is fault-tolerant with an efficient decoder. This construction advances us one step closer towards fault-tolerant random codes.  Practically, these codes also work well for error models that are dynamical in time.

Thus, the dynamic codes we construct in this paper present a new class of quantum error correcting codes and suggest a new route towards universal fault-tolerant schemes for quantum computation, that rely on neither stabilizer codes, nor subsystem codes, nor Floquet codes generated from the gauge group of subsystem codes.

The rest of our paper is organized as follows. In section~\ref{2D_CSS}, we introduce the two-dimensional CSS honeycomb code, discuss it in detail and explain its error-correction properties. In section~\ref{3D_fracton}, we elaborate on an example that generalizes CSS honeycomb codes to three dimensions and show that the instantaneous code cycles between different realizations of the checkerboard and X-cube model. In section~\ref{tree_codes}, dynamic tree codes are introduced and argued to be a more general structure (that need not be periodic) bridging the honeycomb code and the CSS honeycomb codes. We propose a PFA construction of error-correcting protocols and also generalize dynamic tree codes to 3D.

\section{2D CSS honeycomb code}
\label{2D_CSS}

We propose a dynamic quantum error correcting code built solely out of $X$ and $Z$-flavored check operators -- for this reason, we refer to this code as the CSS honeycomb code.  Recall that in the honeycomb code of Hastings and Haah \cite{hastings2021dynamically}, one picks a 3-colorable planar graph and assigns labels of $X$, $Y$, and $Z$ to each of the three orientations of the edges.  The edges of the graph are also three-colorable, and the dynamic measurement protocol consists of measuring the two-body Pauli operators (``checks'') of the flavor corresponding to the orientation of the bond at all the edges of a given color at each round. The color of the edge is defined by the colors of the two plaquettes that it connects, see Fig.~\ref{2D_CSS}.

\begin{figure}[t]
\vspace{0pt}
\centering
\vspace{0pt}
\includegraphics[width= 0.85\columnwidth]{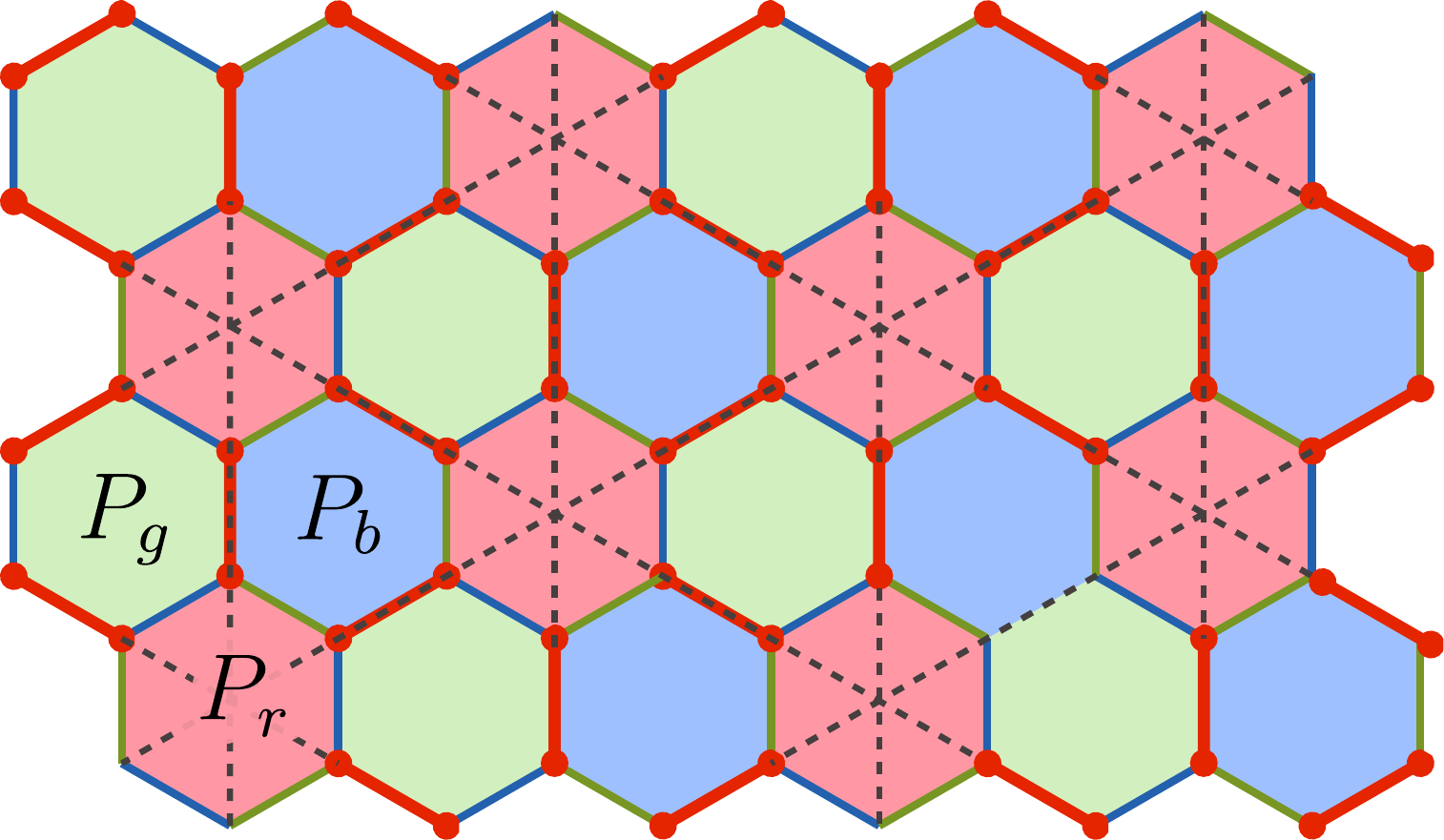}
	\caption{ Fragment of a honeycomb lattice with three-colored plaquettes ($P_{r,g,b}$) and edges. The red, blue and green checks correspond to the edges connecting two plaquettes of the same color. The red checks ($\rc{r}$) which are measured in rounds $3n$ are shown by bold lines and the triangular superlattice is shown by dashed black lines.  }
	\label{fig:CSS_FC}
\end{figure}

\begin{table*}[t]
\begin{tabular}{|c|c|c|c|c|c|c|c|c|c|}
\hline
\multirow{2}{*}{$r$}&\multicolumn{4}{c|}{ISG  $\mathcal{S}(r)$}& \multirow{2}{*}{Syndrome}& \multicolumn{3}{c|}{Logical string}& \multirow{2}{*}{Code}\\
\cline{2-5}\cline{7-9}
& Measure &$m_1$ & $e$ & $m_2$ &  &$m_1$ & $e$ & $m_2$ & \\\hline
-3 & \rXX & &&&&&&&\\
-2 & \gZZ & \PbX &&&&&&&\\
-1 &  \bXX & \PrZ &  \PbX&&&&&&
\\
0 & \rZZ & \PgX & \PrZ & \PbX&
&\gZZ&\rXX&\bZZ&TC(\Rr)
\\
1 & \gXX & \PbZ & \PgX & \PrZ &   \PbX &
\bXX& \gZZ&\rXX&$\overline{\text{TC}}$(\Gg)
\\
2 & \bZZ & \PrX & \PbZ & \PgX &  \PrZ & 
\rZZ &\bXX& \gZZ& TC(\Bb)
\\
3& \rXX & \PgZ & \PrX & \PbZ & \PgX & 
\gXX & \rZZ& \bXX& $\overline{\text{TC}}(\rc{r})$
\\
4 & \gZZ & \PbX& \PgZ & \PrX & \PbZ  &
\bZZ & \gXX & \rZZ& TC(\Gg) \\
5 &  \bXX & \PrZ &  \PbX& \PgZ & \PrX 
&\rXX &\bZZ & \gXX & $\overline{\text{TC}}$(\Bb)
\\
6&  \rZZ & \PgX & \PrZ & \PbX&\PgZ&
\gZZ&\rXX&\bZZ&TC(\Rr)
\\
7 & \multirow{2}{*}{\vdots}&\multirow{2}{*}{\vdots}&\multirow{2}{*}{\vdots}&\multirow{2}{*}{\vdots}&\multirow{2}{*}{\vdots}&\multirow{2}{*}{\vdots}&\multirow{2}{*}{\vdots}&\multirow{2}{*}{\vdots}&\multirow{2}{*}{\vdots}\\
\vdots &&&&&&&&&\\
\hline
\end{tabular}
\caption{Summary of the the CSS honeycomb code. The  table features the measurement sequence, the instantaneous stabilizer group $\mathcal{S}(r)$ at each round, the syndrome plaquettes, logical operators, and the instantaneous codes. The checks and plaquette stabilizers are color-coded for convenience. The `syndrome' column contains the plaquette stabilizers that have been known in previous rounds but are also contained in the checks of the current round. These measurements are used as syndromes for error detection (see Sec.~\ref{sec:decoders}).
The logical operators labeled as electric ($e$) and magnetic ($m_{1,2}$) strings correspond to string operators that violate  the superlattice vertex or plaquette stabilizers of the embedded toric code, respectively.  The magnetic $m_1$ and $m_2$ strings are equivalent up to local operators  acting at their ends. The connection between the logical operators of the CSS honeycomb code and the topological excitations of the embedded codes are explored in Sec.~\ref{sec:TC}.  TC($c$) with $c \in (r,g,b)$ stands for a toric code realized on a triangular superlattice  with  vertices of the superlattice located on plaquettes of color $c$, while $\overline{\text{TC}}$ is the same code conjugated by a layer of single-qubit Hadamards, i.e. where stabilizers have flavors exchanged, $X \leftrightarrow Z$.  }
\label{table:css_floquet}
\end{table*}

In the CSS honeycomb code, the protocol is somewhat simpler and is shown in Table~\ref{table:css_floquet}.  It is partially inspired by the construction of toric code topological order in \cite{Balasubramanian, Balasubramanian2}.  We similarly consider a honeycomb lattice with periodic boundary conditions and divide the plaquettes and the edges into three colors, red, green and blue. At each round of measurements, we apply either red, green, or blue checks. However, the flavor of the check operators applied at each round alternates between $X$ and $Z$ (i.e. one measures two-qubit operators $XX$ or $ZZ$ on the edges of the color of the given round).  This gives a measurement schedule whereby we measure the sequence $\{rXX, gZZ, bXX, rZZ, gXX, bZZ\}$ periodically in time.

Let us start in an arbitrary initial state (alternatively, one prepare a specific state in order to encode logical information in a code) and start measuring checks according to the proposed protocol.  At each round $r$, the state prepared this way is a stabilizer state under an instantaneous stabilizer group (ISG) $\mathcal{S}(r)$. The generators of instantaneous stabilizer groups at each round are listed in Table~\ref{table:css_floquet}. As a remark, similarly to the honeycomb code, instead of post-selecting or correcting to the $+1$ values of the measured stabilizers, we instead record these signs and assume a convention where the ground state is eigenstate of the plaquette stabilizers with eigenvalues determined by the measured signs.

At initial round $r=-3$, the red checks shown in Fig.~\ref{fig:CSS_FC}, which we denote $rXX$, are measured.  At the next round, $r=-2$, we measure $ZZ$-checks on green edges, which anticommute with the measurements at the previous round. However, at this step, the ISG contains the stabilizers $P_b(X)$, which corresponds to a product of Pauli-$X$   around blue plaquettes, and belongs to the center of the group generated by $\la rXX, gZZ \ra$, i.e. commutes with checks of both rounds. 
Measuring $bXX$ in the subsequent round $r=-1$ produces plaquette stabilizers that are the center of the group $\la bXX, gZZ, P_b(X) \ra$, which is $P_b(X)$ and  $P_r(Z)$.  

After measuring $rZZ$ at round $r=0$, the ISG  includes $P_g(X)$, as well as $P_b(X)$ and  $P_r(Z)$ from the previous rounds, as well as current checks  $rZZ$. 
The prepared code has a number of stabilizers that matches the number of qubits on a torus minus two, because the plaquette operators are not all independent.  
This instantaneous code is equivalent to the toric code (TC($r$) in the table). To see this, consider the superlattice formed by the dashed black lines in Figure~\ref{fig:CSS_FC}.  On this triangular superlattice, there are two qubits per edge.  Constraining to the subspace where the checks $rZZ$ simply fuse the two qubits into a single qubit degree of freedom, with effective qubits located on each red edge which have effective logical operators $\Xeff = XX$ and $\Zeff = ZI=IZ$.  Then, it can be seen that $P_g(X)$ and $P_b(X)$ simplify to products of three $\Xeff$ operators around the triangles of the superlattice.  Similarly, $P_r(Z)$ corresponds to the product of $\Zeff$ on the star of the edges emanating out of each vertex of the triangular lattice.  For the simplicity of the presentation, assume that all measured signs of $rZZ$ checks are $+1$ (otherwise, the signs would appear as prefactors in each term in the Hamiltonian without changing the conclusions). Thus, the effective stabilizer code corresponds to the Hamiltonian
\begin{equation}
    H_0^\text{eff}= \sum_{v}A_v(\Zeff) + \sum_{\triangle}B_\triangle(\Xeff)
\end{equation}
where $A_v$ and $B_\triangle$ are the star and plaquette terms on the triangular lattice, respectively.  This Hamiltonian simply describes the toric code, exhibiting the paradigmatic $\mathbb{Z}_2$ topological order.

When we continue implementing the protocol further, the number of logical qubits does not change, and the embedded code in each round is a different realization of the toric code; the period of the embedded code is 6. The logical information is preserved during this cycling, the details of which we address in the next section.  To see that the embedded code changes each round, consider the subsequent $r=1$ step when $gXX$ checks are measured. The value of the stabilizer $P_b(X)$ from the previous step is already contained in the values of the measured green checks, and therefore we do not add it to the list of generators of the instantaneous stabilizer group (ISG) (we add it to the table as a syndrome, however, because the stabilizer value inferred from the green checks at the current round can be compared with the one stored earlier). Additionally, measuring $gXX$ turns the $rZZ$ checks of the previous round to $P_b(Z)$, so the number of logical qubits in the new code does not change.  
We can see that on round $r=1$ we also obtain an effective toric code by drawing a triangular lattice centered on the green plaquettes, and viewing the $gXX$ checks as a fusion of the two qubits on each green edge, which have effective logical operators $\Xeff =XI=IX$ and $\Zeff= ZZ$.  The Hamiltonian corresponding to the embedded code is
\begin{equation}
    H_1^\text{eff}= \sum_{v}A_v(\Xeff) + \sum_{\triangle}B_\triangle(\Zeff),
\end{equation}
which is again a triangular lattice toric code.  

On the next step, $bZZ$ checks are measured, and the plaquette $P_r(Z)$ becomes redundant, so we do not list it in the ISG. A new plaquette $P_r(X)$ is added to the ISG, and the ISG yields an embeddded toric code centered on the blue sublattice (TC($b$)).  
The instantaneous stabilizer groups of the next three rounds are identical to the previous three apart from  $X \leftrightarrow Z$ (and therefore TC code goes into $\overline{\text{TC}}$, see Table~\ref{table:css_floquet}); therefore, the period of the code is 6.

\begin{figure}[t]
\vspace{0pt}
\centering
\vspace{0pt}
\includegraphics[width= 0.65\columnwidth]{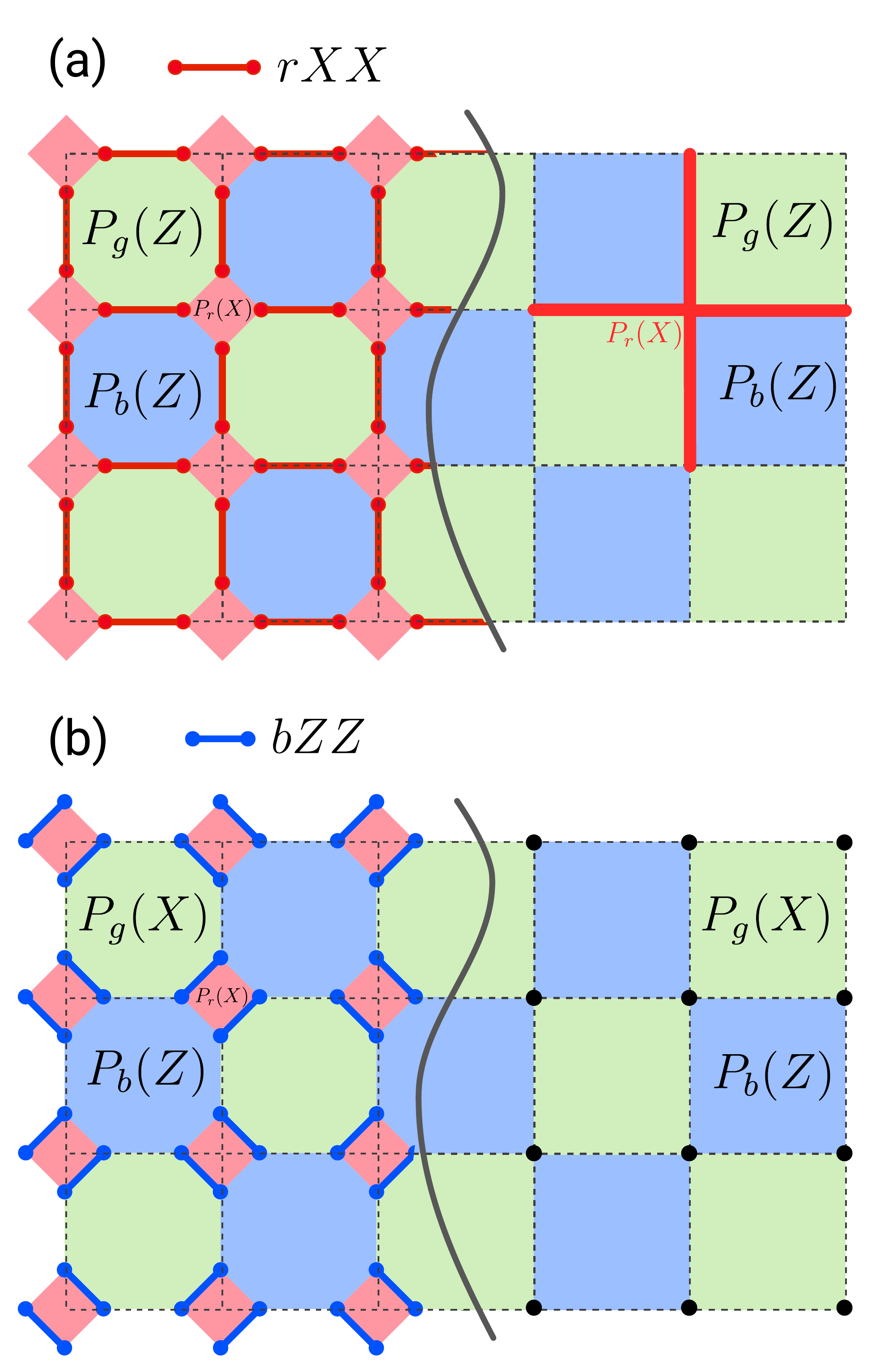}
	\caption{The $rXX$ (a) and $bZZ$ (b) rounds of the CSS honeycomb code realized on the three-colorable square-octagon lattice (periodic boundary conditions are assumed and only part of the lattice is shown for convenience).
	Because the algebraic relations between the checks are the same, and the square-octagon lattice is trivalent with even sided plaquettes, the properties of the square-octagon Floquet code and its error correction are the same as the honeycomb version.  The left half of each lattice shows the original lattice and the ISG, and the right half shows the superlattice. At the $rXX$ step shown in (a), if the two-body checks define local [[2,1,1]] codes, the embedded code on the superlattice is the toric code with qubits on the edges, where $P_{g,b}(Z)$ become the plaquette terms and $P_r(X)$ becomes the star term. In (b) (at blue, and similarly, at green steps), one can view the blue checks together with $P_r(X)$ plaquettes as stabilizers of a [[4,1,2]] local code. This results in a Wen plaquette model where the effective qubits are located on vertices of the square superlattice.
	}
	\label{fig:wen}
\end{figure}

Thus, starting from round $r=0$, our CSS honeycomb code always embeds a toric code in its instantaneous stabilizer group.  A striking difference between the honeycomb code and the CSS honeycomb code is that while the honeycomb code features fixed plaquette stabilizers after three rounds of measurements, the plaquette stabilizers in the CSS honeycomb code change from round to round via substitutions where $P(Z)$ is replaced by another $P(X)$ or vice versa.  This suggests a fundamental difference between our code and the honeycomb code from the perspective of subsystem codes, which we discuss below. In Appendix A, we also show that this code has a regular representation as the same protocol where only $ZZ$-checks are measured at each round and a layer of single-qubit Hadamard gates is inserted after each round. This immediately turns it into a period-3 protocol. Formulated this way using only $ZZ$-checks and unitary layers, the honeycomb code requires single-qubit $S$ and $H$-gates with a period-3 pattern.

Finally, this protocol does not necessarily require a honeycomb lattice and will work on any three-colorable graph, similarly to the honeycomb code \cite{vuillot}. In particular, if we apply the same protocol to the three-colorable square-octagon lattice, the embedded code will alternate between explicitly realizing the Wen plaquette model \cite{PhysRevLett.90.016803}and the toric code on a square superlattice, as shown in Fig.~\ref{fig:wen}.

\subsection{Relation to subsystem codes}

As previously mentioned, subsystem codes are defined by a gauge group $\mathcal G$ that is generically a non-Abelian subgroup of the Pauli group. The stabilizer group of a subsystem code is given by $\mathcal S = \mathcal C(\mathcal G) \bigcap \mathcal G$, where $\mathcal C(\mathcal G)$ denotes the center of the gauge group.  The subsystem code can be viewed as a generalization of the concept of a stabilizer code; the logical qubits of the stabilizer code defined by $\mathcal S$ above are now split into logical qubits and gauge qubits of the subsystem code.  While the logical qubits of the subsystem code are stabilized by $\mathcal G \setminus \mathcal S$, the gauge qubits of the subsystem code are not, and $\mathcal G$ transforms them nontrivially.  Logical operators of $\mathcal S$ are similarly split into `bare' logical operators which act trivially on the gauge qubits, and `dressed' logical operators, which transform the gauge qubits.  Not only do subsystem codes require lower-weight measurements in general (given the non-Abelian nature of the gauge group), but they provide attractive proposals for universal quantum computing.  In particular, one can perform a procedure called gauge fixing, whereby an Abelian subset of the gauge group generators is measured, thus fixing the states of some of the gauge qubits and adding additional stabilizers to $\mathcal{S}$, turning it into $\mathcal{S}'$.  In this way, one can easily switch between different codes with different stabilizer groups, which is called code switching.  

Therefore, one way to construct Floquet codes might come from starting with a subsystem code (`parent subsystem code') and measuring subsets of its gauge operators sequentially, arriving at different effective stabilizer codes as a result. The honeycomb code fits in this framework: the $XX$, $YY$, and $ZZ$ checks of the honeycomb code correspond to the Hamiltonian terms of the Kitaev honeycomb model and generate a subsystem code. Even though this subsystem code does not contain any logical qubits, the dynamical protocol leads to an ISG that, at every round, is the same as that of the parent subsystem code minus two operators that cannot be obtained by such sequential measurements (the `inner' logical operators). In contrast, an attempt to find a subsystem code that would play the same role for the CSS honeycomb code fails, as we show below. 
Note that the concept of the `parent subsystem code'  introduced above is distinct from and unrelated to the `parent code' for anyon condensation, which has been used to independently derive the CSS honeycomb code in ref.~\cite{brown_2022}.

The construction of the CSS honeycomb code does not involve the checks of a subsystem code, and therefore one might ask whether there exists a relation between this dynamical code and any subsystem code. Let us explore this question in more detail.  Consider the group generated by all checks of our protocol, i.e. 
\begin{equation}\label{eq:G}
\mathcal{G}= \left\langle rXX, gZZ, bXX, rZZ, gXX, bZZ\right\rangle.
\end{equation}
For this subsystem code, $\mathcal{S}= Z(\mathcal{G}) = \left \langle \prod X, \prod Z \right \rangle$. These extensive operators contain the $\mathbb{Z}_2$ global symmetry  of the effective codes realized by the Floquet protocol. At each step of the Floquet codes, one of these symmetries is just a product of all checks, and the other one is the product of one color of plaquettes with the opposite flavor of the checks; thus, both operators are contained in the ISG. The latter operator is a symmetry of the embedded toric code.    

Let us show that the subsystem code defined in eq.~\eqref{eq:G} provides only very limited information about the stabilizer codes realized by the Floquet protocols constructed from its checks, and does not play a useful role as a parent subsystem code. Assume that we gauge fix the code $\mathcal{G}$ by adding the checks $rXX$ to the stabilizer group. The new code is $\mathcal{G}' = \left\langle bXX, rZZ, gXX \right\rangle$ and $\mathcal{S}' = \left \langle \prod X, \prod Z, rXX \right \rangle$. This code does not bear resemblance to the topological codes realized by the Floquet protocol, and such topological codes cannot be achieved by further gauge fixing or removing some of the gauge checks. Therefore, even though the CSS honeycomb code is generated by sequentially measuring the checks of the subsystem codein \eqref{eq:G}, this subsystem code does not construe a useful parent subsystem code, unlike in the case of the honeycomb code.   

We may also introduce a concept of \emph{$k$-sliding subsystem code}, which is defined by a gauge group  $\mathcal{G}_k^r$ generated by subset of checks from $k$ consequent rounds  $r-k+1, \dots, r$. Let us see if this relaxed notion of subsystem code can be a parent subsystem code for the Floquet code at some of the rounds. First, we notice that if $k = 1$, the $k$-sliding subsystem code stabilizes too many qubits. Before proceeding to higher $k$, we note that without loss of generality, we can consider round $r = 6n$:
\begin{itemize}
    \item $k = 2$: The generators of the gauge group are simply 
    \begin{equation}
        \mathcal{G}_k^r = \mathcal{G}_2^{6n}= \left\langle bXX, rZZ, \prod X, \prod Z \right\rangle.
    \end{equation}
    The center of this gauge group is $Z(\mathcal{G}_2^{6n}) = \left\langle P_g(Z), P_g(X), \prod X, \prod Z \right\rangle$, which does not have three distinct types of plaquette stabilizers.  
    \item $k = 3$: 
    \begin{equation}
        \mathcal{G}_3^{6n}= \left\langle gZZ, bXX, rZZ\right\rangle.
    \end{equation}
The center of this gauge group is $Z(\mathcal{G}_3^{6n}) = \left\langle P_g(Z), P_r(Z),  s(rZZ-gZZ), \prod X, \prod Z  \right\rangle$, where $s(rZZ-bZZ)$ are strings of $rZZ$ and $gZZ$ checks along homologically nontrivial cycles of the torus. Moreover, since the stabilizers are of the same flavor, the code is classical. As we see, the stabilizer group  also does not contain three distinct types of plaquette stabilizers.  
    \item $k \geq 4$:  A similar exercise shows that the stabilizer group of a 4-sliding subsystem code produces a single flavor of a plaquette as well as the global symmetries. There is a more fundamental reason that looking beyond to $k \geq 4$ is unreasonable.  Recall that during each round, a $c=r,g$ or $b$-colored plaquette substitution occurs where $P_c(X)$ is replaced with $P_c(Z)$.  In the next round, the value of $P_c(X)$ is then destroyed (and the same happens on rounds where $X \leftrightarrow Z$).  Given that after 4 rounds the values of a stabilizer is destroyed by measurements, there is no reason one should consider a subsystem code formed by checks from too many subsequent rounds. 
\end{itemize}
Therefore, even though the 2-sliding subsystem code contains more stabilizers because it `inherits' additional stabilizers from two previous measurement rounds, this nevertheless tells us that there is no useful concept of a parent subsystem code for the CSS honeycomb code, even if we generalize to a $k$-sliding subsystem code. This discussion also indicates that the CSS honeycomb code might belong to a different class of dynamic codes than the honeycomb code. Another example of a Floquet code that does not seem to have an immediate  parent subsystem code is the automorphism code~\cite{aasen}, although at the time of writing, these codes have not been shown to be error correcting or fault tolerant.

As we show below, our code nevertheless conserves logical information and possesses a threshold. Surprisingly, this shows that subsystem codes are not necessary for the construction of `good' error-correcting dynamic codes.

\begin{figure}[t]
\vspace{0pt}
\centering
\vspace{0pt}
\includegraphics[width= 0.85\columnwidth]{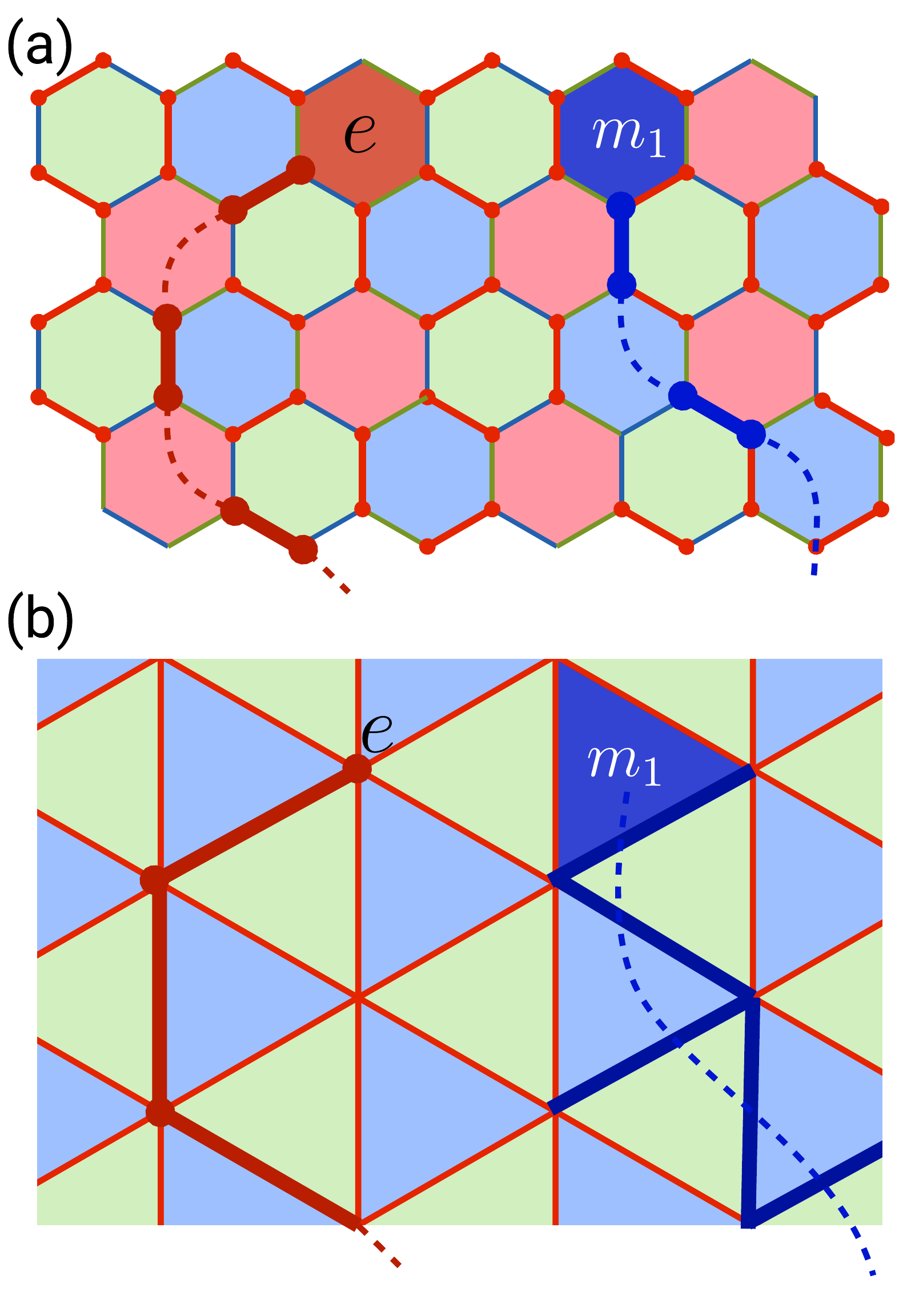}
	\caption{String operators generating anyon excitations $e$ and $m_1$ anyons shown on (a) honeycomb lattice and (b) the superlattice (with qubits on the edges), occurring at rounds $r = 3n$. At steps corresponding to odd $n$, the $e$-string is formed from $rXX$ checks, whereas $m_1$($m_2$)-strings are generated by $gZZ$($bZZ$) check strings, respectively, as shown in the figure. On the triangular superlattice,  the red plaquettes turn into vertices whereas the blue and green plaquettes correspond to the two types of triangular plaquettes. At rounds corresponding to even $n$, the picture is the same upon exchanging $X \leftrightarrow Z$. 
	}
	\label{fig:anyons}
\end{figure}

\subsection{Conservation of logical information and automorphism}

\label{sec:TC}

The instantaneous  code embedded in our dynamic code is equivalent to the toric code, as indicated in the last column of Table~\ref{table:css_floquet}.  As mentioned before, the three plaquette stabilizers of the ISG become the stabilizers of a toric code on a triangular superlattice (see Figs.~\ref{fig:CSS_FC}and \ref{fig:anyons}). String operators, whose endpoints anticommute with the vertex or plaquette stabilizers, excite $e$ and $m$-anyons on their ends, respectively. Note that on the triangular lattice, there are two types of $m$-anyons ($m_{1,2}$), corresponding to two orientations of triangular plaquettes, though they nevertheless belong in the same superselection sector. These strings for $e$ and $m_{1,2}$ particles are formed by checks indicated in Table~\ref{table:css_floquet}, see Fig.~\ref{fig:anyons}.  For example, at round $6n$, the string operators creating $e$ and $m$ particles are
\begin{equation}
    S_{e,i}= \prod_{\ell \in \mathcal{P}_{i,r}}(XX)_{\ell}\hspace{0.6cm}S_{m_{1(2)},j}= \prod_{\ell \in \mathcal{P}_{j,b(g)}}(ZZ)_{\ell}
\end{equation}
where $\mathcal{P}_{i,c}$ is a set of checks of color $c$ that forms a string emanating from hexagon $i$ of the same color.  The two $X$ and two $Z$ logical operators for the instantaneous code can be obtained by taking $e$ and $m_{1,2}$-type strings around homologically nontrivial cycles of the torus. The $m_{1}$-type string taken along such a cycle is the same as $m_{2}$-type string; hence, they both produce the same logical operator. 

Table~\ref{table:css_floquet}shows the $m_1$ anyon at each step turns into the $e$ anyon of the subsequent round, and the $e$ anyon turns into the $m_2$ anyon. The $m_2$ anyon `disappears', which occurs exactly when the syndrome for plaquettes violated by this flavor of anyon is measured. In fact, the $m_2$ anyon is equivalent to $m_1$ one up to the checks of the current round, therefore the information carried in magnetic logical operator is not lost.  This will be also useful for us when understanding the error correcting properties of this code.

Similarly to the transformation of anyons, the $X$ and $Z$-type logical operators swap at each step but are never measured; thus, this code conserves logical information.

In the honeycomb code, the logical operators can be classified as either `inner' or `outer' ones. The `inner' operators are products of checks along the homologically nontrivial cycles of the torus and they belong to the stabilizer of the honeycomb subsystem code, whereas the `outer' ones do not. 
Because of the lack of a subsystem code framework, there is no concept of `inner' and `outer' logical operators as in the original honeycomb code. 

Finally, we remark that it might appear as though our code does not possess an $e \leftrightarrow m$ automorphism (which the currently existing examples \cite{hastings2021dynamically,aasen}of Floquet codes do). 
However, let us follow a magnetic string $gZZ$ measured in round 0 (see Table~\ref{table:css_floquet}). It is preserved for three steps, and at step 2 the magnetic logical produced by $gZZ$ and $rZZ$ is the same. Now, following $rZZ$ to round $3$, we see that it becomes an electric string. The codes at step 0 and 3 are the same toric code conjugated by on-site Hadamard transformations, and therefore, up to this transformation, an automorphism has in fact been performed.

\subsection{Decoders and threshold}
\label{sec:decoders}

Despite the absence of the overarching structure of a subsystem code and a stationary ISG, the CSS honeycomb code surprisingly possesses a threshold. For a simplified $X,Z$-error model of single-qubit errors, we can reduce the decoding problem to that of the honeycomb model~\cite{hastings2021dynamically}, and thus, argue that our code has a threshold.  For other error models and more specialized decoders, the thresholds for these two codes are likely quantitatively different.

In the simplified error model, we only  consider occurrence of single-qubit $X$ and $Z$ errors with probability $p$, corresponding to the quantum channel $\mathcal{E}(\rho) = (1-p)^2 I \rho I + p(1-p) X \rho X + p(1-p) Z \rho Z + p^2 Y \rho Y$. Since $X$($Z$)-type single-qubit error can be commuted past similarly-flavored checks, we only need to consider the occurrence of an  $X$($Z$)-type error after even(odd) rounds. We simplify our error model assuming independent single-qubit errors of $X$($Z$)-type. Because the error syndromes for errors occurring after even(odd) rounds are measured on odd(even) timestamps only, error correction can be performed separately on even and odd temporal sublattices. $X$ and $Z$ errors are treated identically, so for simplicity we will deal with $Z$ errors in what follows.

\begin{figure}[t]
\vspace{0pt}
\centering
\vspace{0pt}
\includegraphics[width= 0.85\columnwidth]{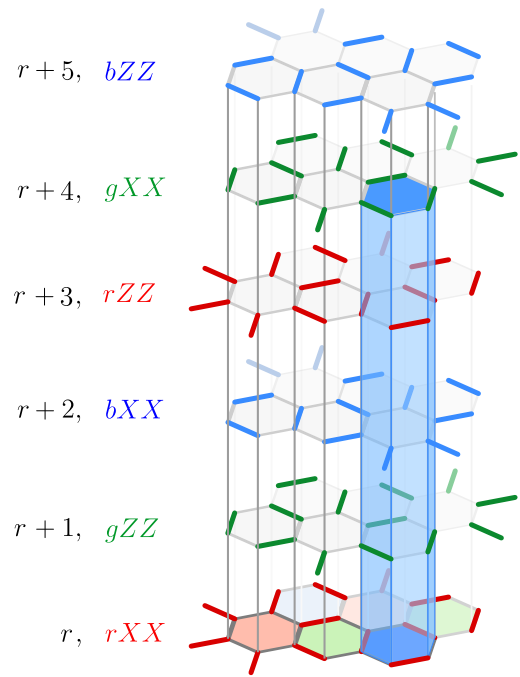}
	\caption{
	A $P_b(X)$-type detector cell  is shaded blue; it can be violated by $Z$-type errors occurring on its spacetime support (not including $t=r+4$). It consists of measuring the highlighted $P_g(X)$ plaquette at time $r$ and then measuring it again at time $r+4$. Two such neighboring detection cells are required to determine the spacetime location of the edge where the single-qubit error has occurred.
	}
	\label{fig:detector}
\end{figure}

We consider a simple non-optimal decoder for this error model.  The value of each type of plaquette stabilizer is measured twice during each period-6 cycle, once at step $r$, and once at step $r+4$ (before being erased at step $r+5$), as shown in Fig.~\ref{fig:detector}. Comparing these values allows one to infer the error syndromes that are necessary for decoding. Thus, one needs to record the syndromes inferred from comparing the the two values of each plaquette stabilizer obtained at different rounds.

For each type of the plaquette stabilizer, there exists a type of check that anticommutes with this stabilizer (for example, consider the pair $P_r(X)$ and $rZZ$). Thus, the values of such plaquette stabilizers are necessarily randomized after enough time, and it is impossible to fix the values of stabilizers in our Floquet code, say, to $+1$, once and for all. The only constraint on the values of the stabilizers in the system is that the logical information is preserved, which was argued above. Despite this, it is surprising that the protocol is still error-correcting.

In more detail, each plaquette stabilizer, once inferred from check measurements, survives for 4 rounds before a check is applied that anticommutes with this stabilizer. This allows us to define a corresponding spacetime detector cell. For example, consider the plaquette $P_b(X)$ that is inferred at step $r$  from $rXX$ checks ($r = 6n+3 (\text{mod}\ 6)$ in Table~\ref{table:css_floquet}). One such plaquette at round $r$ is highlighted  in Fig.~\ref{fig:detector}. Once measured, this plaquette is not randomized for the next three rounds, and appears in the ISG. At round $r+4$, this plaquette is measured again (the `syndrome' plaquette in Table~\ref{table:css_floquet}). 
The respective spacetime detector cell is supported on this plaquette between the rounds $r$ and $r+4$, i.e. between  two consecutive measurements of $P_b(X)$ occurring at these rounds.
The product of these two plaquette measurements determines the error syndrome and therefore whether the detector cell has been violated.
As an example, the product $\{P_b(X)\}_{r}\times  \{ P_b(X)\}_{r+4}$ will be equal $+1$ if there is an even number of $Z$-type errors in its support and $-1$ if there is an odd number of errors. 
Finally, at round $r + 5$, $P_b(X)$ plaquettes are randomized by $bZZ$ checks.

If a $Z$-error has occurred after an odd timestep, it can be commuted past the measurement of the next round, and therefore it is sufficient to only consider $Z$($X$)-type errors after even(odd) rounds.   Consider first an isolated single-qubit $Z$-type error occurring after an even round of the protocol. First of all, knowing only the edge where the error has occurred is sufficient to correct the error. In this situation, correction corresponds to applying a Pauli operator with the same flavor as the error to any of the qubits on this edge; if the qubit was guessed incorrectly, this turns the single-qubit error into a check error of the current round. However, one can show that such check errors do not affect the logical state in the instantaneous toric code, so long as the edge is of the same color as the checks of the last round.  

Now we discuss how to determine the edge where the error occurs.  Let us assume that a $Z$ error occurs after round $r=0$. At the next round $r=1$, measurement of one of the $gXX$ checks involving this qubit will acquire an error. This will violate a $P_b(X)$-type detector cell (its type is determined by the `syndrome' column in Table~\ref{table:css_floquet}). Similarly, at round $r=3$,  the $rXX$ checks are measured, and one of the checks changes its sign due to the error. This violates a respective $P_g(X)$-type detector cell. These two detector cells share only one $ZZ$-type edge: the red edge at $r=0$. Thus, the spacetime location of the faulty edge caused by a single-qubit error is determined unambiguously. Finally, in the case of multiple errors, a minimum-weight perfect matching decoder  can be used (discussed below) and the location of error chains will be determined up to stabilizers of the code.

The principle of error correction after this point is same as in the honeycomb code, where the violated detectors have to be matched on the same spacetime lattice (decoding graph).  This spacetime lattice parameterized by$\{\bm b_1, \bm b_2, \bm b_3\}$, where $\bm b_1 = \lp \frac{\sqrt{3}}{2},\frac{3}{2},2 \rp$ and the other two vectors are obtained by $\frac{2\pi}{3}$ rotation of each previous one around the time axis (the distance between the centers of hexagons is taken to be $\sqrt{3}$, see Fig.~\ref{fig:CSS_FC}).
 The links of the lattice are given by $\{\bm b_1, \bm b_2, \bm b_3\}$. We have two copies of the syndrome lattice,  corresponding to odd and even timesteps, that store syndromes from  bit-flip ($X$) and phase-flip ($Z$) errors, respectively. 

 The minimum-weight perfect matching decoder can be applied for error correction  \cite{dennis2002topological}and the problem can be similarly mapped to a random-bond Ising model with a phase transition to a confined ``non-correctible'' phase, thus exhibiting a threshold. Thus, the CSS honeycomb code has a threshold similarly to the honeycomb code.

Another way of convincing oneself of a threshold comes from the probability of error leading to a failure being exponentially suppressed as $L \rightarrow \infty$ for a torus of size $L \times L$. Let us sketch a bound on the failure probability in a similar manner to  ref.~\cite{dennis2002topological}. Each edge on the syndrome lattice has a one-to-one correspondence with a physical qubit of a toric code on the superlattice from one timestep earlier. Therefore, once the minimum-weight perfect matching decoder determines the lowest weight string of errors $E_0$ by finding the shortest string on the syndrome lattice, the set of links on the respective superlattices at the times when errors in $E_0$ occur will be determined unambiguously. As we noted earlier, only the edge of the superlattice on which the error occured needs to be detected (i.e. the errors have to be detected up to a position within the two-qubit check of the round after which the error occured).  Because of this one-to-one correspondence, we can refer to the `flipped' edges on the syndrome lattice as to errors. 
Let $p$ be the probability of a single-qubit error and let us assume that there are no measurement outcome errors. We consider one type of error, and thus, one copy of the syndrome lattice, and assume that the true error string is $E$ and that the recovery string is $E_0$ and has length $w_0$.

Notice that the set $C = E + E_0$ contains a set of disjoint loops, either homologically trivial or not.   A failure occurs if the set $C$ contains at least one homologically nontrivial cycle. Consider an arbitrary connected path $S(w)$ of length $w$, then the probability of failure can be bounded by the probability of $C$ containing a path $S(w)$ with length greater than the distance of the code:
\be
\mathbb{P}(\text{fail}) \le \sum_{S(w): w \ge d = L}\mathbb P (S (w) \subseteq C).
\ee
We consider only self-avoiding walks, since every closed loop can be eliminated. Assume that we have an arbitrary path containing $w_e$ errors (i.e. links belonging to $S(w) \bigcap E)$). The probability of such a path is $\binom{w}{w_e }p^{w_e}(1-p)^{w-w_e}$.  Now, if the path $S(w)$ happens to be contained in $C$,  the number  of errors on it $w_e > \frac{w}{2}$ because of the assumption of the minimum weight matching. We formulate this as $S \setminus \lp S \bigcap E \rp \subseteq E_0$. This yields:
\begin{align}
   \binom{w}{w_e }p^{w_e}(1-p)^{w-w_e}\le \binom{w}{w_e }p^{\frac{w}{2}}.
\end{align}
Thus, we have
\be
\begin{split}\label{eq:prob_string}
\mathbb{P}&(S(w) \subseteq  C) \le \sum_{S(w): S \setminus \lp S \bigcap E \rp \subseteq E_0}\mathbb{P}(S (w) ) \\
&\le \sum_{w_e \ge \frac{w}{2}}\binom{w}{w_e }p^{\frac{w}{2}}\le 2^{w}p^{\frac{w}{2}}.
\end{split}
\ee
The probability of failure can be bounded by:
\be
\begin {split}
\mathbb{P}(\text{fail}) \le \sum_{S(w): w \ge d = L}2^{w}p^{\frac{w}{2}}=  \sum_{w \ge L}n_S(w) 2^{w}p^{\frac{w}{2}},
\end{split}
\ee
where $n_S(w)$ is the total number of  self-avoiding walks on the syndrome lattice. The lattice has 6 directions for the walk from each point, and therefore $n_S(w) \le  6 \times 5^{w}\times \frac{1}{9}L^2 T$, where $\frac{1}{9}L^2 T$ counts the number of the possible starting points. 
Therefore:
\be
\begin {split}
\mathbb{P}(\text{fail}) \le   \frac{10}{3}L^2 T  \sum_{w \ge L}5^{w}2^w p^{\frac{w}{2}}\le \frac{10}{3}L^2 T \frac{\left( 10^2 p \right) ^{\frac{L}{2}}}{\lp1-10\sqrt{p}\rp }.
\end{split}
\ee
The probability of the failure that we found above is exponentially suppressed as long as $p,q < p_c^{(0)}= 0.01$ and the timescale of running the code $T(L)$ before the error correction is performed satisfies $\lim_{L \rightarrow \infty}L^2 T \left( 10^2 p \right) ^{\frac{L}{2}}= 0$ for given $p<p_c^{(0)}$ (which is always the case, for example, for $T(L) = \text{poly}(L)$). Thus, the lower bound on the threshold within this model is $p_c \ge p_c^{(0)}$.

A threshold in the $X,Z$-error model implies a threshold against measurement outcome errors as well, because a check error corresponds to a correlated-in-time application of a non-commuting single-qubit Pauli error right before and after the check is applied. 
Additionally, a partial implementation of any logical operator is correctable in a similar sense to how the `inner' logical operators are correctable in the honeycomb code. Despite the fact that there is no concept of an `inner' logical operator because the subsystem code structure is absent in our code, any partially implemented logical operator is detectable. 
Robust error correction during rounds $r=-3$ to $r=0$ is not possible because the instantaneous code is still being prepared during these steps.  One can instead start by initializing the effective toric code at $r=0$ on the corresponding superlattice by a different high-fidelity method. Similarly, the measurement of the logical operators can be done by two-qubit checks applied to the effective toric code on the superlattice after termination of the protocol.

In the independent work\cite{brown_2022}, the CSS honeycomb code has been discovered independently, and the threshold of $
\sim 0.3 \%$ has been found numerically for MWPM decoder for depolarizing noise model (at the circuit level). We refer the reader to this reference for a useful discussion of error correction for this code and the effect of measurement errors.

\section{3D generalization: Fracton Floquet code}
\label{3D_fracton}

In this section, we present an example of a 3D construction inspired by our 2D CSS honeycomb  code, which we find gives rise to Floquet codes for fracton topological orders.

\begin{table*}[t]
\begin{tabular}{|c|c|c|c|c|c|c|c|c|c|}
\hline
\multirow{2}{*}{$r$}&\multicolumn{4}{c|}{ISG}& \multirow{2}{*}{Syndrome}& \multicolumn{3}{c|}{Logical string}& \multirow{2}{*}{Code}\\
\cline{2-5}\cline{7-9}
& Measure &$m_1$ & $e$ & $m_2$ &  &$m_1$ & $e$ & $m_2$ & \\\hline
-3 & \rXXX & &&&&&&&\\
-2 & \gZZ & \cubebX&&&&&&&\\
-1 &  \bXXX & \cubebZ &  \cubebX&&&&&&\\
0 & \bZZZ & \sqgX & \cubebZ & \cubebX&&\gZZ&\rXXX&\rZZZ&CB(\Bb)\\
1 & \gXX & \cuberZ & \sqgX &  \cubebZ &   \cubebX & \bXXX& \gZZ&\rXXX&$\overline{\text{XC}(\gc{g})}$\\
2 & \rZZZ & \cuberX & \cuberZ & \sqgX &  \cubebZ & \bZZZ &\bXXX& \gZZ& CB(\Rr)\\
3& \rXXX & \sqgZ & \cuberX & \cuberZ & \sqgX & \gXX & \bZZZ& \bXXX& $\overline{\text{CB}(\rc{r})}$\\
4 & \gZZ & \cubebX & \sqgZ & \cuberX & \cuberZ  &\rZZZ & \gXX & \bZZZ& XC(\Gg) \\
5 &  \bXXX & \cubebZ &  \cubebX& \sqgZ & \cuberX &\rXXX &\rZZZ & \gXX & $\overline{\text{CB}(\bc{b})}$\\
6&  \bZZZ & \sqgX & \cubebZ & \cubebX&\sqgZ&\gZZ&\rXXX&\rZZZ&CB(\Bb)\\
7 & \multirow{2}{*}{\vdots}&\multirow{2}{*}{\vdots}&\multirow{2}{*}{\vdots}&\multirow{2}{*}{\vdots}&\multirow{2}{*}{\vdots}&\multirow{2}{*}{\vdots}&\multirow{2}{*}{\vdots}&\multirow{2}{*}{\vdots}&\multirow{2}{*}{\vdots}\\
8&&&&&&&&&\\
\hline
\end{tabular}
\caption{Summary of the the three-dimensional CSS fracton Floquet code. The table features the measurement sequence, the instantaneous stabilizer group $\mathcal{S}(r)$ at each round, the syndrome plaquettes, logical string operators, and the instantaneous codes. The `syndrome' column stores the star/volume stabilizers that can be inferred from the checks of the current round and compared with known value of this plaquette stabilizer in the previous round. Strings of checks are labeled electric ($e$) and magnetic ($m_{1,2}$) in correspondence with the instantaneous code on the superlattice.  The magnetic $m_1$ and $m_2$ strings are equivalent up to local operators at their ends. XC($g$) stands for the embedded X-cube model realized on a cubic superlattice with effective qubits on its edges. CB($r/b$) stands for the embedded checkerboard model realized on a cubic superlattice with effective qubits on its vertices and the volume stabilizers  defined on $r/b$ cubes.  $\overline{\text{XC}}$ ($\overline{\text{CB}}$) are the same codes conjugated by a layer of single-qubit Hadamards, i.e. having stabilizers exchange flavors $X \leftrightarrow Z$.  }
\label{table:fractons}
\end{table*}

\begin{figure}[t]
\vspace{0pt}
\centering
\vspace{0pt}
\includegraphics[width= 1\columnwidth]{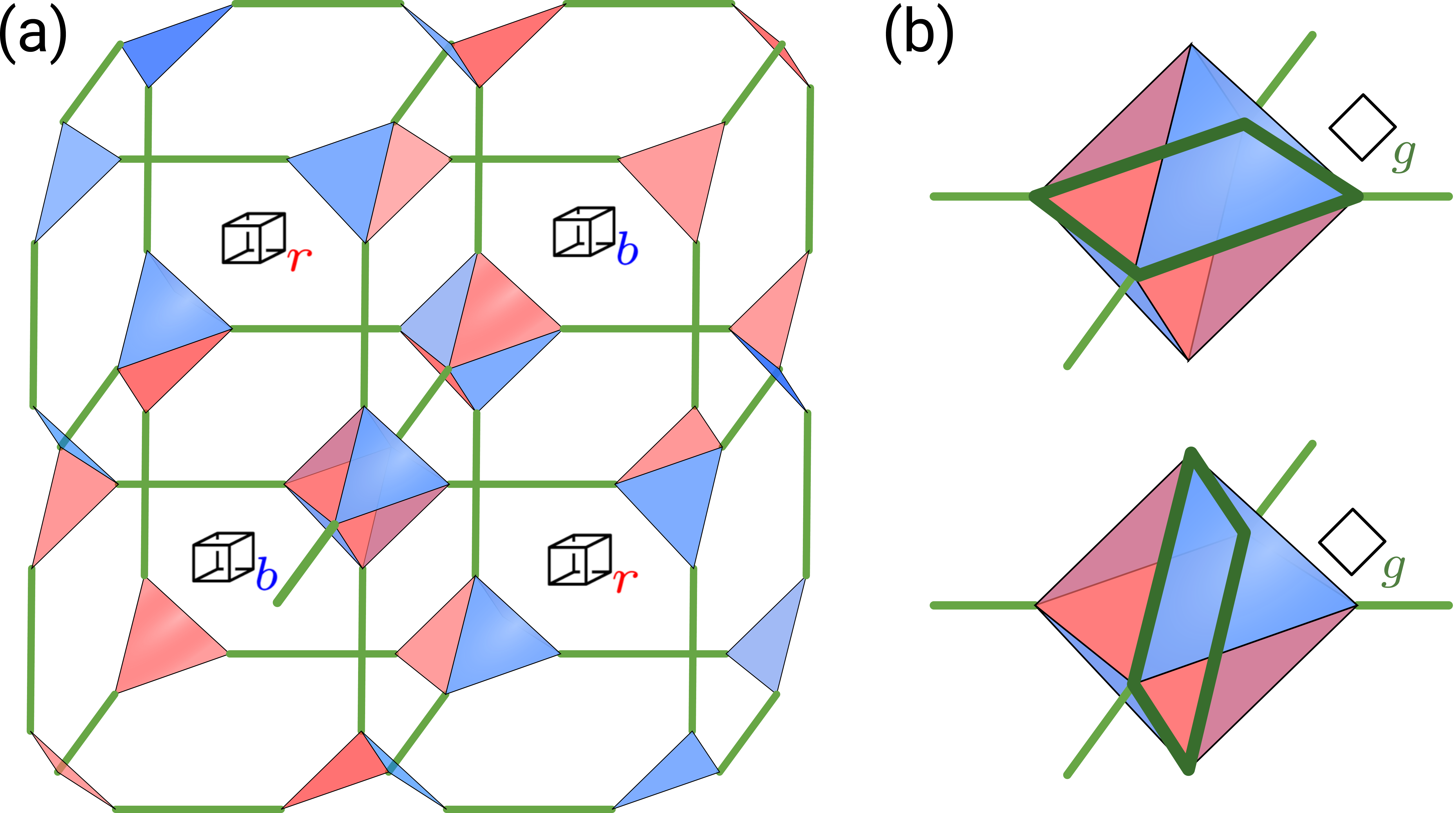}
\caption{(a) Decorated cubic lattice with two qubits per edge (located at the vertices of the resulting lattice). The cubes of the two types correspond to \mancube$_\rc{r}$ and \mancube$_\bc{b}$, respectively, and the triangular plaquettes between the octahedra located at each vertex and the cube of type $\bc b$($\rc r$) are shaded red(blue), respectively, i.e. the complementary color. The two  square plaquettes $\diamond_\gc{g}$ that produce two independent stabilizers are shown in (b).}
\label{fig:fracton}
\end{figure}

The general protocol and associated geometry are shown in Figure~\ref{fig:fracton}and Table~\ref{table:fractons}.  In particular, we consider a truncated cubic lattice, which can either be thought of as a cubic lattice, where every site is turned into an octahedron, or a lattice of corner-sharing octahedra where each shared corner is extended into an edge. The physical qubits are located at the vertices of this lattice. It can be seen that the volumes of this lattice are three-colorable: we label the cubic volumes  with red and blue (\mancube$_\rc{r}$ and \mancube$_\bc{b}$ in Figure~\ref{fig:fracton}) and the octahedra with green. 

The protocol is implemented using checks of  weight two and three. The checks that we will use in the protocol consist of the products of three Paulis $rXXX$($rZZZ$) around red triangles, $bXXX$($bZZZ$) around blue triangles, and the two-body checks along green links $gXX$($gZZ$), see Fig.~\ref{fig:fracton}. This coloring was chosen to match the coloring in the 2D CSS honeycomb code: the green edges protrude out of green volumes, and red(blue) plaquettes interface between the volumes of two other colors (blue/green and red/green, respectively).

We repeat the measurement procedure similar to the 2D case. The protocol is outlined in Table~\ref{table:fractons}.  After round -2 when the green links are measured, the product of the eight red triangular plaquettes that form the truncated faces of the blue cube are in the instantaneous stabilizer group.  This is notated by a volume stabilizer
\begin{equation}
  \mbox{\mancube}_b (X) = \prod_{\triangle \in \mbox{\mancube}_b}(rXXX)_\triangle.
\end{equation}
In the next round $r=-1$ we measure the products of $X$s on the blue triangles, indicated by $bXXX$. These do not commute with the green checks of the previous round, but instead commute with the product of green checks around a blue cube.  Hence, only the product of $gZZ$ checks forming $\mbox{\mancube}_{b}(Z)$ stays in the instantaneous stabilizer group.

Naively, at round 0 one might wish to return to the beginning of the protocol and measure $rZZZ$,  in accordance with the 2D protocol for the CSS honeycomb code.  However, the $rZZZ$ check commutes with $bXXX$ of the previous step, which results in 6 independent plaquette stabilizers on a single octahedron.  Since an octahedron is comprised of 6 physical qubits, this stabilizes a single state.
The fix is to instead measure $bZZZ$ at round 0, thereby cycling between the check colors in the sequence ``$rgb\,bgr \ \, rgb\,bgr \ldots$'' rather than ``$rgb \ rgb \ldots$''.  Measuring $bZZZ$ preserves both $\mbox{\mancube}_{b}(Z)$ and $\mbox{\mancube}_{b}(X)$, but does not commute with $bXXX$ of the previous round.  Instead, on each octahedron, the square  plaquettes $\diamond_g (X)$ remain in the instantaneous stabilizer group of the next round, which we refer to as `diamond' stabilizers.  There are three such diamonds in total, but only two of them are independent.  Using the fact that the product of $bZZZ$ around an octahedron is the identity, there are five independent stabilizers per octahedron which means that a single qubit per octahedron (or, equivalently, a single qubit per vertex of the cubic lattice) effectively remains.  The equivalent Hamiltonian for all of the stabilizers at round 0 is therefore
\begin{equation}
    H_0 = J\sum_{\octahedron}\lp bZZZ + \diamond_g(X) \rp + \sum_{\mbox{\mancube}_b}(\mbox{\mancube}_b (X) + \mbox{\mancube}_b (Z)).
\end{equation}
For $J \gg 1$, the first term contains five independent stabilizers that fuse the six qubits of the octahedron into one effective qubit with local effective Pauli operators acting on it being $\Xeff = rXXX$ and $\Zeff=rZZZ$. Now, since $\mbox{\mancube}_b (X)$($\mbox{\mancube}_b (Z)$)  are comprised of products of $rXXX$($rZZZ$), each of which now acts as an effective Pauli on each vertex, the operators reduce to a product of $\Xeff$($\Zeff$) around the eight vertices of the blue cubes
\begin{align}
        H_0^\text{eff}&= \sum_{\mbox{\mancube}_b}(\mbox{\mancube}_b (X) + \mbox{\mancube}_b (Z)), \end{align}
\begin{align}
         \mbox{\mancube}_b (X) &= \prod_{v\in \mbox{\mancube}}\Xeff_v,\end{align}
\begin{align}      \mbox{\mancube}_b (Z) &= \prod_{v\in \mbox{\mancube}}\Zeff_v.
\end{align}
This is nothing but the \emph{checkerboard model}~\cite{VijayHaahFu16}defined on the cubic superlattice. 

At round 1, we measure $gXX$ on the green links.  This set of measurements includes the $\mbox{\mancube}_{b}(X)$ stabilizer of the previous round (which is therefore updated and stored for determining the syndrome, see Table~\ref{table:fractons}), and also adds a new stabilizer $\mbox{\mancube}_{r}(Z)$ formed from the checks of the previous round.  The Hamiltonian formed from the instantaneous stabilizer group is then
\begin{equation}
    H_1 =  J \sum_{\ell}gXX_{\ell}+   \sum_{\octahedron}\diamond_g(X) +  \sum_{\mbox{\mancube}_{r,b}}\mbox{\mancube}(Z).
\end{equation}
For $J \gg 1$, the $gXX$ checks fuse the two qubits on each green edge of the cubic lattice into a single qubit per edge with effective Pauli operators $\Xeff = XI =IX$ and $\Zeff = ZZ$. The effective model is
\begin{align}
        H_1^\text{eff}&=  \sum_{\octahedron}\diamond_g(X) +  \sum_{\mbox{\mancube}_{r,b}}\mbox{\mancube}(Z).\\
         \diamond_g(X) &= \prod_{e\in \diamond}\Xeff_e,\\
          \mbox{\mancube}(Z) &= \prod_{e\in \mbox{\mancube}}\Zeff_e.
\end{align}
Namely, the $\diamond_g(X)$ stabilizers are the star stabilizers and the $\mbox{\mancube}(Z)$ are the cube stabilizers of the \emph{X-cube model}~\cite{VijayHaahFu16}.  There might exist a link between the emergence of the X-cube model subsequent to the checkerboard model and the fact that two coupled copies of an X-cube model are connected to the checkerboard model by an adiabatic deformation~\cite{PhysRevB.99.115123}.

At round 2, one measures $rZZZ$, after which the checks of the previous round form the stabilizer $\mbox{\mancube}_r (X)$ and the stabilizer $\mbox{\mancube}_b (Z)$ is contained in the newly measured checks (and will thus be used for determining the syndrome).  The effective Hamiltonian is then
\begin{equation}
    H_2 = J \sum_{\octahedron}\lp  rZZZ + \diamond_g (X) \rp + \sum_{\mbox{\mancube}_r}(\mbox{\mancube}_r (X)+\mbox{\mancube}_r (Z)),
\end{equation}
which is again the checkerboard model, but now on the red cubes.

This concludes a period's worth of measurements and upon repeating the measurement sequence a similar cycling continues. To summarize, the embedded code alternates between two types of type-I fracton: the checkerboard model centered on $b$($r$) cubic sublattice and the X-cube model. Additionally, an $X \leftrightarrow Z$ mapping occurs every round, and the period of the code is 6.

We note in passing that the current protocol can be modified  by measuring a periodic sequence that alternates between $(rXXX, \  gXX)$ and $(bZZZ, \  gZZ)$. This increases  the rank of the ISG and fuses the two qubits on each edge of the cubic lattice into one effective qubit at each round.  This  protocol is equivalent to repetitive measurements of the three-body $X$ and $Z$ check operators of the subsystem toric code proposed in ref.~\cite{kubica2022single}.

\subsection{Relation to subsystem codes}

The conclusion about the relation between our 3D construction and subsystem codes is the same as in 2D. First of all, the stabilizer of the gauge group generated by all checks in the protocol contains only the subsystem symmetries shared by all ISGs of the Floquet code (i.e. the subsystem symmetries shared by the checkerboard and X-cube fracton orders).   These are products of $X$ and $Z$ operators on planes formed by green checks. At each round, these operators are either contained in the last measured checks or in the product of one of the types of the volume stabilizers. Some of these operators  become `inactive' logicals that we discuss in next section. 

Similarly to the CSS honeycomb code, gauge fixing the subsystem code comprised of all checks does not provide any useful information for construction of 3D Floquet code.    

Consider further the following gauge groups for the $k$-sliding subsystem codes (noting that $k = 1$ is trivial):
\begin{itemize}
    \item $k = 2$: The relevant gauge group is
    \begin{equation}
        \mathcal{G}_2= \left\langle rXXX, gZZ\right\rangle.
    \end{equation}
    The center of this gauge group $Z(\mathcal{G}_2)$ contains $\left\langle \text{\cuberZ{}, \cubebX}, \prod_{planes}X, \prod_{planes}Z\right\rangle$ along with  a possible sub-extensive number of string-like/plane-like operators.   
    \item $k = 3$: The relevant gauge group is
    \begin{equation}
        \mathcal{G}_3 = \left\langle rXXX, gZZ, bXXX\right\rangle.
    \end{equation}
    The  local stabilizers contained in $Z(\mathcal{G}_3)$ are $\left\langle  \text{\cuberX, \cubebX}, \prod_{\text{planes}}X, \prod_{\text{planes}}Z\right\rangle$ along with a subextensive number of string-like/plane-like operators again.
    \item It is clear that the center will be a group that is not larger than that of $Z(\mathcal{G}_3)$ upon further adding checks to the gauge group at $k \geq 4$.
\end{itemize}
Therefore, we again conclude that there is no single sliding subsystem code whose stabilizer group contains the set of plaquettes of any ISG for the 3D Floquet code.

\subsection{Conservation of logical information}

Consider the decorated cubic lattice on a $T^3$ torus of size $2L_x\times 2L_y \times 2L_z$, where the even-sized linear dimensions are required for three-colorability. The effective X-cube model on the corresponding superlattice has a ground state degeneracy of $4(L_x+L_y+L_z)-3$ while that of the effective checkerboard model is $4(L_x+L_y+L_z)-6$. Thus, there seems to be a discrepancy in the number of logical operators in the corresponding rounds. The resolution to this puzzle is a feature not present in the 2D code; there are three logical operators of the static X-cube model that are read out or scrambled by the measurement schedule, and therefore do not belong to the set of logical qubits in the Floquet code. We will call such logical qubits and respective operators \emph{inactive} logical qubits and operators.  In contrast, the remaining $4(L_x+L_y+L_z)-6$  logical qubits of the static code that store information in the Floquet code will be called \textit{active}. In fact, the inactive logical operators that are read out/measured are among the symmetries in the center of the subsystem code $\mathcal G$. 

To see what happens explicitly,  we first start from round one ($r=1$ mod 6), where the ISG corresponds to the X-cube model. Let us recall how to count the logical operators for the corresponding instantaneous effective code. Given a straight line along the effective cubic lattice, the product of $\Zeff_e$ on all edges along the line commutes with the X-cube stabilizers. This physically corresponds to tunneling a lineon excitation around the torus. Moreover,  $\Zeff_e$ strings applied along different parallel lines are distinct, since they are not related by a product of stabilizers. However, there is a relation between certain products of such logical operators. Taking a product of four adjacent parallel lines that form edges of a cube is equal to a product of the enclosed $\mbox{\mancube}(Z)$ stabilizers. For concreteness, let us pick logical operators formed by products along the lines in the $z$-direction. There are $(2L_x)(2L_y)$ such lines. 
There are also  $(2L_x-1)(2L_y-1)$ relations imposed on these lines, one for each square in the $xy$ plane minus conditions that product of all the cubes in a plane equals identity. All together, the $z$-lineons give rise to $(2L_x)(2L_y)-(2L_x-1)(2L_y-1) =2L_x +2L_y -1$ independent $Z$ logicals\footnote{These are $Z$-logicals of $\overline{\text{XC}(g)}$, but would be called $X$-logicals of $\text{XC}(g)$.}.  Summing over the other two directions, we find $4(L_x+ L_y + L_z) -3$ logical $Z$ operators.

\begin{figure}[t]
\vspace{0pt}
\centering
\vspace{0pt}
\includegraphics[width= 1\columnwidth]{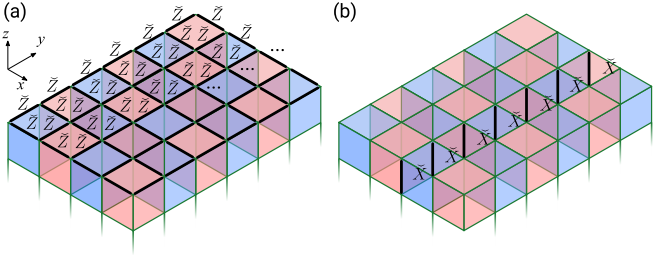}
\caption{A fragment of the cubic superlattice of rounds $r=1$~mod~6 with realization of the X-cube model $\overline{\text{XC}(g)}$. The qubits are fused by the $gXX$ checks of rounds $r=1$~mod~6 into a single effective qubit per edge. On a 3-torus, examples of the inactive logical $Z$ and $X$ operators are shown in panels (a) and (b), respectively. There are total of three such independent operators for each cycle around the torus.  }
\label{fig:inactive}
\end{figure}

Now, consider the logical operators formed by a product of all $\Zeff =gZZ$ along all edges in a fixed $xy$ plane. Measuring $rZZZ$ in the next round $r=2$, we note that such product of $gZZ$ in the plane is equivalent to the product of all $\diamond_g(Z)$ in the same plane, see example in  Fig.~\ref{fig:inactive}. Moreover, each diamond is a product of two $rZZZ$ operators. It therefore follows that this particular logical operator of the X-cube model is measured in the next round $r=2$, and is therefore inactive. Similarly, the product of $gZZ$ along all edges in one fixed $xz$ and one fixed $yz$ plane is also measured. This accounts for the three inactive $Z$ logicals.

Next, let us  similarly find the active $X$-logical operators, i.e., those that commute with the measurements of the next round $r=2$. Define a product of $rXXX$ along a straight line. Suppose the line points in the $z$ direction, which tunnels a lineon, which is a bound state of an $xz$-planon and a $yz$-planon\footnote{It can also be thought of as a bound state of two fractons that share a diagonal edge.}. Like the $Z$-type lineons, there is a similar constraint on the $X$-type lineons: the product of tunneling four adjacent lineons forming the edges of a cube can be decomposed into a product of stabilizers, and the product of tunneling all lineons for a fixed plane is the identity. The importance of defining the bound states is that the local hopping operators come in pairs, and hence they will commute with $rZZZ$ checks of the next round, which is the neccessary condition for them being active. This gives the total of $4(L_x+ L_y + L_z) -6$ active $X$ logicals. Now we ask what the remaining three logical operators are which anticommute with the checks of round two. Fixing a direction, say $y$, consider the product of $\Xeff$'s  that hops a planon (which is a bound state of two fractons) across the $y$ direction, as shown in Fig.~\ref{fig:inactive}. It is clear that this operator anticommutes with one of the $rZZZ$ operators. Moreover, this operator is unique up to stabilizers and the active $X$-logical operators. Lastly, it anticommutes the inactive $Z$ logical operators in the $xy$ plane. By rotational symmetry, we conclude that there are three such inactive $X$ logicals. 

Finally, let us confirm that the active logical qubits indeed survive and are transferred to logical qubits  of round 2, which is the checkerboard model. The active $X$ logical operators are products of $bXXX$, which is the effective qubit $\Xeff$ of round $r=2$. They therefore transfer faithfully to the $X$ logical operators of the checkerboard. As for the active $Z$ logical operators defined as products of $gZZ$ along a line, using the $rZZZ$ checks  of round two, we find that their strings are equivalent to strings of $bZZZ$ operators:
\begin{equation}
\nonumber
\includegraphics[width=\columnwidth]{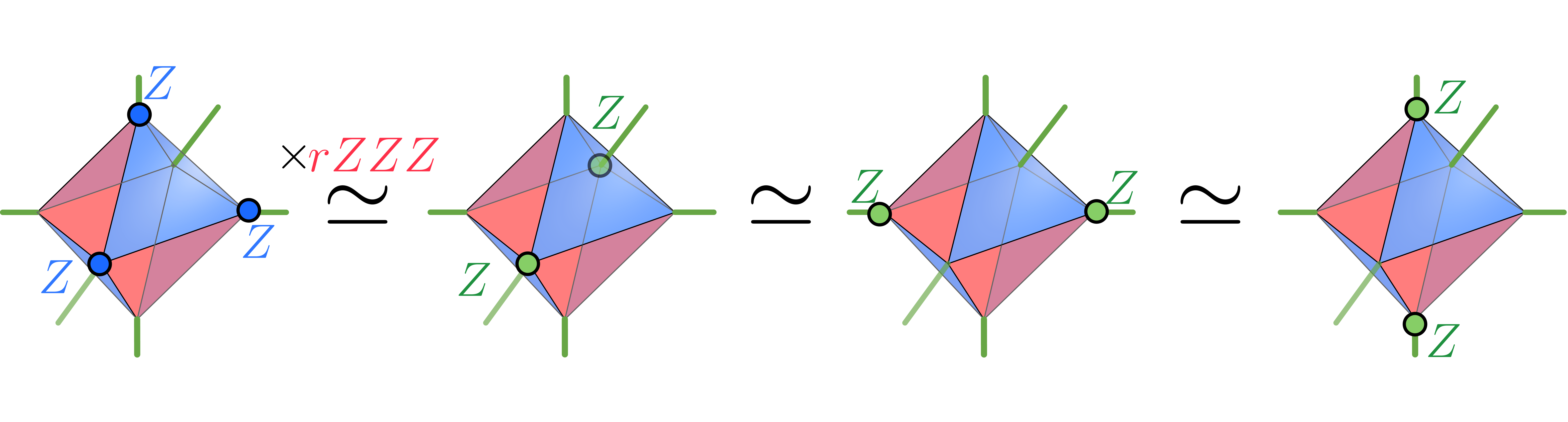}
\end{equation}
which is the product of  effective local $\Zeff$ of round two. They therefore faithfully transport into the logical operators of the checkerboard model.

Going from round $r=2$ to $r=3$, we see that the ISG is defined on the same lattice. We find that the $Z(X)$  logical operators of CB($r$) in round $r=2$ become $Z(X)$ logical operators of $\overline{\text{CB}(r)}$, which equate to $X(Z)$ logical operators of CB($r$) in round three. We hence conclude that the $X$ and $Z$ logical operators are swapped.

Finally from round $r=3$ to $r=4$, the logical information is transferred from a product of $bXXX$ to a product of $gXX$ using the measurements $rXXX$:
\begin{equation}
\nonumber
\includegraphics[width=\columnwidth]{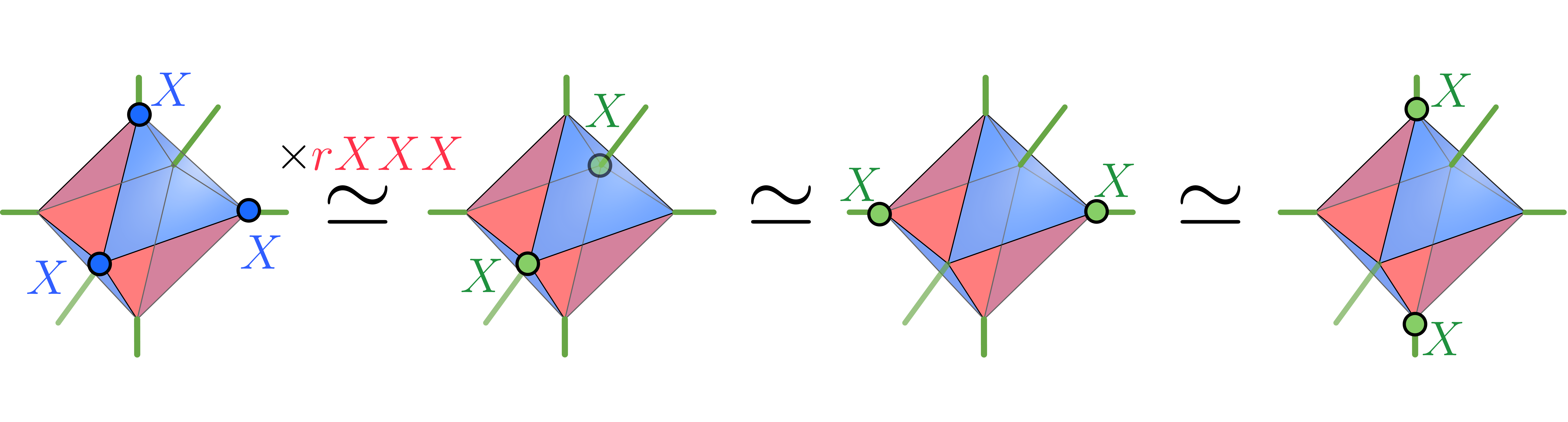}
\end{equation}
Thus, the logical operators of the checkerboard model embed into the active logical operators of X-cube in the next step. The next three rounds, $r=5,6,7$, proceed identically but with $X$ and $Z$ swapped.

Finally we discuss the automorphism occurring in the model. The obvious one that exchanges between magnetic and electric sectors is seen from comparing CB($r$) and $\overline{\text{CB}(r)}$ in rounds 2 to 3 and comparing $\overline{\text{CB}(b)}$  and CB($b$) in rounds 5 to 6, where the role of the $X$ and $Z$ logical operators are swapped. Therefore, the automorphism between these codes occurs in the same sense as in 2D Floquet code, up to a layer of Hadamard gates. Comparing $\overline{\text{XC}(g)}$ and XC($g$)
in rounds 1 and 4 is more subtle. Although the active logical operators get swapped,  this does not produce a well-defined automorphism for the X-cube model. The inactive logical operators cannot be permuted, since they have different spatial support. Therefore, transfer of the logical information from electric to magnetic sectors given by the protocol does not preserve the fusion rules for the excitations. It would be interesting to examine  more generally the connection between the existence of inactive logical qubits and the absence of an automorphism in Floquet codes during certain rounds.

\subsection{Decoders and threshold}
 
The error correction in 3D Floquet fracton codes is remarkably similar to that in 2D CSS honeycomb codes, mainly because the former is a natural generalization of the latter. The details of the error syndromes looks different, as we discuss in more detail below, but the decoder  on (3+1)-dimensional spacetime lattice that is a  generalization of the (2+1) dimensional case considered earlier will perform well, and will  have a threshold by an analogous argument. Moreover, if analyzed using statistical mechanics mappings~\cite{dennis2002topological,https://doi.org/10.48550/arxiv.2112.05122}, larger thresholds are likely to be facilitated thanks to the higher dimensionality. 

Considering the same error model as before, where $X$ and $Z$ errors occur with probability $p$, we only need to consider two distinct times for the errors to occur, and the rest of the behavior of syndromes can be deduced by symmetry.  We also find that the syndromes for the errors that occurred after even(odd) rounds are measured at odd(even) rounds only. This means that the errors after even and odd timesteps can be corrected separately. 

Consider first a $Z$ error occurring at a single qubit right after round 0. One of the $gXX$ checks of round 1 will be affected, as well as  two cubes \cubebX \ inferred using this check, which can then be compared with the stored value and recorded as a syndrome. This allows one to determine the green link on which the error occurred. Then, on step 3, $rXXX$ is measured, and two triangles on the same octahedron will have their values flipped. The three (redundant)  \sqgX \ plaquettes belonging to the same octahedron can be inferred using by combining two triangles belonging to this octahedron. These stabilizers can be compared with the values that have been stored earlier, and the comparison uniquely determines the diagonal of the octahedron on which the fault occurs. Together with knowing the green link where the error occurs, this allows one to unambiguously determine the location of the error.  However, the same syndrome is found when the error instead occurs on two complementary qubits belonging to the same blue triangle. In this case one can still assume a (more probable) single-qubit error and correct for it. If the actual error occurred on the two complementary qubits, the total error will become a $bZZZ$ operator, which is inconsequential, because it corresponds to the check of the round $r=0$ after which the error occurred. Similarly to the 2D case, these kinds of check errors do not affect the logical state.

Errors occurring after round $3n+2$ lead to a syndrome that is qualitatively similar to the one discussed above. A  qualitatively different type of error syndrome is found for errors that occur after rounds $r=3n+1$. Without loss of generality, consider an $X$ single-qubit error occurring after round $r=1$. In round $r=2$, two of the $rZZZ$ checks will be flipped and two cubes \cubebZ \ sharing an edge whose values are inferred from these checks will be flipped and stored as a syndrome. Similarly, a pair of cubes \cuberZ will be flipped in $r=4$. The syndromes at both rounds $r=2$ and $r=4$ allow  to determine the location of the flipped edge $\ell$, and the additional redundancy can be used for error correction that is more robust against the measurement outcome errors. Applying a correcting single-qubit Pauli-$X$ to any of the qubits on this edge will either correct the error or apply an $XX$ to the entire edge. In the latter case, this will be removed once the round that re-measures this check occurs. A pair of errors on two neighboring qubits belonging to the same octahedron produces the same syndrome as the pair of errors on the other two qubits belonging to the same  $\diamond_g$-plaquette. Nevertheless, this error can still be corrected up to an inconsequential edge error on a green link  by applying a Pauli operator on any of the qubits not belonging to this diamond. 

Thus, the syndromes will occur on the spacetime lattice that is formed by the centers of cubic volumes of the same color at $t=r$ (mod 3) and $t = r+2$ (mod 3), and by the vertices of the cubic lattice at $t=r+1$  (mod 3). At times $t=r+1$, the measured syndrome can take one of eight values, indicating which of the three $\diamond_g$ square plaquettes have (or have not) been violated.
Mapping this problem onto a graph matching problem and designing an efficient minimum-weight decoder based on the known syndromes is an involved task that we leave to a future work.

\section{Dynamic tree codes}
\label{tree_codes}

\begin{table*}[t]
\begin{tabular}{|c|c|c|c|c|c|c|c|c|}
\hline
\multicolumn{1}{|c|}{\multirow{2}{*}{$r$}}& \multicolumn{4}{c|}{ISG}                                   & \multicolumn{1}{c|}{\multirow{2}{*}{Syndrome}}& \multicolumn{1}{c|}{\multirow{2}{*}{$m_1$}}& \multicolumn{1}{c|}{\multirow{2}{*}{$e$}}& \multicolumn{1}{c|}{\multirow{2}{*}{$m_2$}}\\ \cline{2-5}
\multicolumn{1}{|c|}{}                 & \multicolumn{1}{c|}{Measure}& \multicolumn{3}{c|}{Plaquettes}& \multicolumn{1}{c|}{}& \multicolumn{1}{c|}{} & \multicolumn{1}{c|}{}    & \multicolumn{1}{c|}{}                 \\ \hline
\vdots & \vdots & \vdots & \vdots & \vdots& \multirow{2}{*}{\vdots}& \vdots & \vdots & \vdots
\\
0  & $\rc {r f_0 f_0}$ & \Pg$(f_{1})$   &\Pr$(f_0)$ & \Pb$(f_{-1})$ &  & $\gc {g f_0 f_0}$ & $\rc {r f_1 f_1}$ &  $\bc {b f_0 f_0}$ 
\\
1  & $\gc {g f_1 f_1}$ & \Pb($f_2$) & \Pg($f_1$) & \Pr($f_0$) &  
 \Pb$(f_{-1}) = \left\{\begin{matrix}
&\prod_{\bc \varhexagon}\gc {g f_1 f_1}, \ \ \text{if }f_1 =  f_{-1}
\\ 
&\prod_{\bc \varhexagon}\gc {g f_1 f_1}\prod_{\bc \varhexagon}\rc {r f_0 f_0},\ \ \text{if }f_1 \neq  f_{-1}
\end{matrix}\right.$ 
&  $\bc {b f_1 f_1}$ & $\gc {g f_2 f_2}$ &  $\rc {r f_1 f_1}$  
\\
2  & $\bc {b f_2 f_2}$ & \Pr($f_3$) &  \Pb($f_2$) & \Pg($f_1$) & \vdots &  $\rc {r f_2 f_2}$ &  $\bc {b f_3 f_3}$ &  $\gc {g f_2 f_2}$
\\
3  & $\rc {r f_3 f_3}$ & \Pg($f_4$) &  \Pr($f_3$) & \Pb($f_2$) & \Pg$(f_1) = \left\{\begin{matrix}
&\prod_{\gc \varhexagon}\rc {r f_3 f_3}, \ \ \text{if }f_3 =  f_1
\\ 
&\prod_{\gc \varhexagon}\rc {r f_3 f_3}\prod_{\gc \varhexagon}\bc {b f_2 f_2},\ \ \text{if }f_3 \neq  f_1
\end{matrix}\right.$ & $\gc {g f_3 f_3}$ & $\rc {r f_2 f_2}$ &   $\bc {b f_3 f_3}$\\
\vdots & \vdots & \vdots & \vdots & \vdots& \vdots & \vdots & \vdots & \vdots \\
  \hline
\end{tabular}
\caption{The random-flavor Floquet code, where the checks follow a fixed color sequence $rgb$, but the flavors $f_r$ in each round $r$ are randomized (with the constraint that $f_{r+1}\ne f_r$).
The syndromes that are obtained in rounds 1 and 3 and are listed in the `Syndrome' column. We only show the syndromes obtained in odd rounds which are used to detect the error after even rounds and omit listing the syndromes at even rounds for clarity. 
If we assume that a single-qubit Pauli error of flavor $f_0$ occurred after round $r=0$, the listed syndromes will allow one to unambiguously determine the red edge where the error has occurred. }
\label{table:tree_2D}
\end{table*}

We have considered CSS versions of Floquet codes in both two and three dimensions.  Both these codes have robust error-correcting properties, but fall outside of the subsystem code formalism.  In this section, we further generalize these results by introducing a broader family of dynamic codes where the measurement sequence \emph{need not be}periodic.  Surprisingly, under certain constraints on correlated randomness of the measurements, this random code can correct arbitrary single-qubit Pauli errors. This construction bears relation to some classes of monitored random circuit codes and random unitary circuits~\cite{PhysRevX.9.031009,PhysRevX.10.041020,PhysRevLett.127.235701,PhysRevB.103.174309,PhysRevB.98.205136,PhysRevLett.125.030505,PhysRevB.103.104306,https://doi.org/10.48550/arxiv.2105.13352}, achieving practical quantum error correction in which has been a long-standing challenge~\cite{PhysRevB.103.174309,PhysRevB.98.205136,PhysRevLett.125.030505,PhysRevB.103.104306,https://doi.org/10.48550/arxiv.2105.13352}. 
As of now, it is unclear whether  random circuits, including those considered in refs.~[\onlinecite{ali2022}] and [\onlinecite{https://doi.org/10.48550/arxiv.2207.07096}], consisting of randomly applied checks of the honeycomb code,  can possess  a finite threshold.

We call the proposed random codes \emph{dynamic tree codes}because a given code carves out a path on a configuration space of allowed checks which forms a tree. Dynamic tree codes can be viewed as the first instance  monitored random circuit codes that are capable of correcting arbitrary single-qubit Pauli errors, though they are restricted to correlated randomness and the absence of spatial randomness.  Practically speaking these codes might be useful if the error model itself is dynamical: for example, it could  adapt the error correction procedure to biased error models and to adversarial time-dependent error models.

\subsection{Random-flavor Floquet codes and switching between CSS honeycomb code and honeycomb code}
We start with the 2D case. Let us show that if the colors of the checks follow an $r g b$ sequence  but the Pauli flavors are randomized such that the flavors of two consecutive rounds are different, the resulting \emph{random-flavor Floquet code}will be error-correcting and will have a threshold. The condition on flavors of two consecutive rounds being different ensures that the checks of the two rounds always anticommute and the rank of the ISG stays the same.  

Without loss of generality, we can consider a code shown in Table~\ref{table:tree_2D}(considering only four arbitrary rounds is sufficient for the argument), where $f_r \in \{X,Y,Z \}$ stands for the flavor of the given round $r$, $f_{r+1}\neq f_r$, and the colors of the checks follow $rgbrgb...$  sequence (or the symmetric thereof, $rbgrbg...$).  We can also assume that the code has been properly initialized far in the past.  By inspection, we see that this code realizes a sequence of toric codes by analogy with 2D CSS and honeycomb codes on a superlattice corresponding to the color of the current round; the ISG at each round is shown in the table.  From the logical strings of the code shown in  Table~\ref{table:tree_2D}, we see that the condition $f_{r+1}\neq f_r$ and $c(r+1) \neq c(r)$ is indeed sufficient for  conserving logical information between the rounds, because is ensures that one never measures logical operators from round to round. 

Let us now show that the random-flavor Floquet code can correct arbitrary single-qubit Pauli errors. The error after each round $r$ can be expanded in the basis of Pauli flavors $f_r$ and $f_{r+1}$ of the current and the next rounds, respectively. The $f_{r+1}$ component of the error can be commuted past the checks of round $r+1$, and only the error of the flavor the current round $f_r$ needs to be considered. Therefore, we again need only consider the error model where single-qubit Pauli errors have the flavor of the last round.  

Without loss of generality, we can consider an $f_0$-Pauli error that occurred after round $r=0$. Let us show that we can detect the red edge where this error occurred in spacetime  in our random-flavor Floquet code shown in Table~\ref{table:tree_2D}. This is sufficient for being able to correct the error: as before, we only need to apply the $f_0$ Pauli operator on any of the qubits of the edge. If we guess the wrong qubit, the result is the two-Pauli operator equivalent to the check of the last round which is an inconsequential error.

At round $r=0$, prior to the error, the values of plaquettes $P_g(f_1)$ and $P_b(f_{-1})$ are known. If one re-measures the values plaquettes $P_g(f_1)$ and $P_b(f_{-1})$ after the error has occurred, the change of the sign of these plaquettes will allow to determine the edge where the error has occurred, which in this case is a red edge between these two plaquettes.  Referring to Table~\ref{table:tree_2D}, we see that indeed, the $P_b(f_{-1})$ and $P_g(f_1)$ plaquettes are immediately re-measured at round $r=1$ and $r=3$, respectively, yielding the needed syndrome changes. This allows us to detect and correct single-qubit errors. Additionally, note that the locations and the timestamps of the errors are the same as in the CSS honeycomb code and in the honeycomb code.

\begin{figure*}[t]
\centering
\includegraphics[width= 1\textwidth]{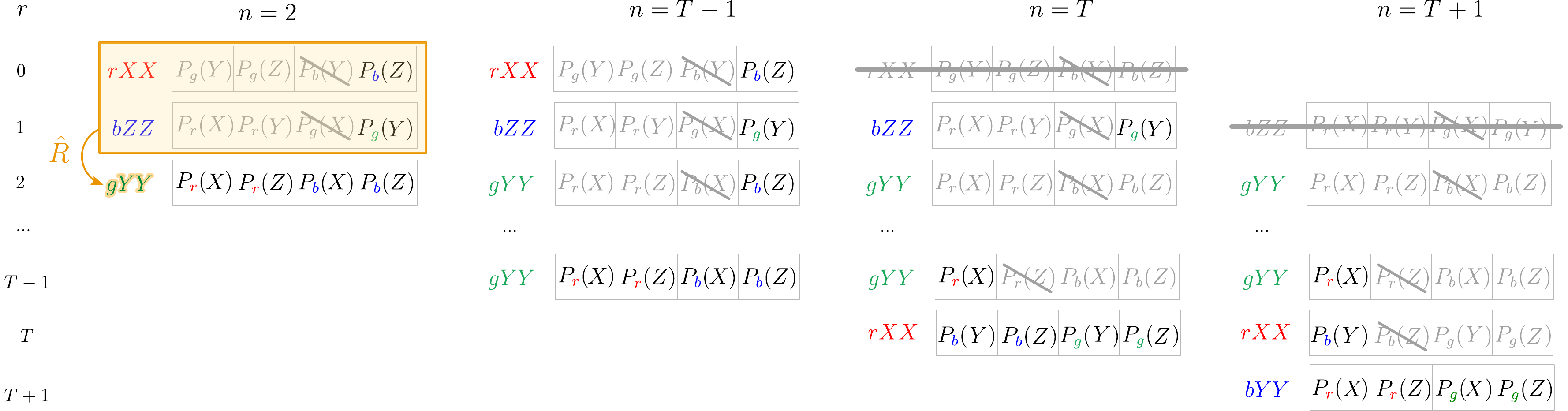}
	\caption{An example of a random sequence of measurements using the $T$-PFA scheme discussed in Sec.~\ref{sec:PFA}.  The PFA stores a running window containing $T$  arrays of cells (apart from during the initialization rounds between $1$ and $T$). We assume that the code has  been properly initialized in a toric code state at $r=0$. Here, $r$ counts the rounds  in the  measurement history and $n$ is the current round in the PFA operation. 
	The arrays correspond to the $T$ memory cells of the $T$-PFA.
	In each array, the darkened plaquette label indicates a plaquette that needs to be remeasured in order to infer the syndrome corresponding to a possible error on the plaquette. 
	Light gray plaquettes that are not crossed out indicate that the plaquette has been already remeasured. We keep the plaquettes that are redundant up to the checks of the current round for completeness.    
	Light gray plaquettes that are crossed out have been randomized before they could have been remeasured. 
	Each new check is chosen by the PFA based on the memory of checks and plaquettes, corresponding to the set of update rules of the PFA denoted by operator $\hat{R}$ described in the main text. 
	In short, the new check is chosen such that it differs in color and flavor from the previous check and such that it does not randomize any of the `not-remeasured' (darkened) plaquettes in the memory.  Furthermore, checks must also be chosen such that the measurement of all syndromes from more than $T$ steps in the past is completed; so long as this holds true, it suffices for the PFA to store memory from the past $T$ rounds and for the error to be undetected only for up to $T$ rounds.}
	\label{fig:PFA}
\end{figure*}

Random-flavor Floquet codes are fault-tolerant by a similar argument to that for the CSS honeycomb and the honeycomb codes. If the ISG and syndrome are book-kept similarly to how it is shown in Table~\ref{table:tree_2D}, it is clear that the plaquette occurring for the first time at round $r$ will be kept in the code until round $r+3$ when it will be updated. The new value of the plaquette will be able to detect errors that have occurred after rounds $r$ and $r+2$. This is the same syndrome-error relation as in the codes that have been studied earlier and the decoding procedure is analogous to that of the honeycomb code. 

One consequence of the existence of error-correcting random-flavor Floquet codes is that one can switch between the protocols for the honeycomb code and the CSS honeycomb code fault-tolerantly, so long as the color sequence of the checks remains unperturbed.  Therefore, this shows that the CSS honeycomb code and the honeycomb code are compatible, which might be useful for future design of error correcting codes with time dependent error models.

\subsection{PFA construction of error-correcting dynamic codes}\label{sec:PFA}

Next, we address the question whether it is possible to introduce more randomness into the Floquet code protocol while preserving its ability to detect and correct errors.  We propose a construction that uses what we call a $T$-probabilistic finite automaton ($T$-PFA). If initialized in a toric code ground state, the automaton chooses the flavor and color of each check at random, realizing a toric code at each round and by design guarantees that any single-qubit Pauli error will be detected no later than $T$ steps after its occurrence and is thus correctable (we use the same error model as earlier in the text). If this protocol is initialized in an arbitrary state, it will prepare the toric code model no later than after $T$ steps, and will continue to function as described, with a length-$T$ window for error detection.

The construction is outlined below and is exemplified in Fig.~\ref{fig:PFA}.   In the discussion below, we assume that the protocol is initialized in a toric code state corresponding to the first check of the protocol. 

The $T$-PFA has memory containing  $T$ arrays, each consisting of up to 4 plaquettes with status either `not remeasured' or `remeasured'. When a new check at round $r=n+1$ is measured, 4 plaquettes with `not remeasured' status are added to the corresponding array. Two of the plaquettes are elements of the current ISG with flavors that are not the same as the flavor $f_{n+1}$ of the current check.  The other two are  equivalent to the first two plaquettes up to checks of the current round. The memory is designed this way because if an error occurs after round $r=n+1$, remeasuring any two of these four plaquettes of different colors would be sufficient to detect the edge where the error has occurred and its timestamp. We keep these plaquettes in memory in order to ensure that the possible syndrome for a single-qubit Pauli error can be tracked and recorded.  

The update rules  for $T$-PFA after $r=n$ are:
\begin{enumerate}
    \item Pick the check of the next round from ($rXX,rYY,rZZ,gXX,gYY,gZZ,bXX,bYY,bZZ$):
    \begin{enumerate}
        \item Eliminate the checks which have the same color or flavor as the past round $r=n-1$ (this guarantees that the rank of the ISG stays the same).
        \item Eliminate checks that randomize plaquettes that are stored in memory with `not remeasured' status.    \item If in memory cell with $r=n-T$ there is a `not remeasured' plaquette, choose only from checks that remeasure this plaquette.
        \item Otherwise, pick a random check from the remaining options.
    \end{enumerate}
    \item Update the memory based on the new check.
    \begin{enumerate}
        \item Scan the memory for `not remeasured' plaquettes. For each such plaquette change the status to `remeasured' if the value of this plaquette can be inferred based on the current check and the interim checks. 
        \item Erase any plaquettes in the memory that are redundant with those already remeasured.
        \item Erase any plaquettes that have been randomized by the new check.
        \item Erase the array at round $r=n-T$ (as the syndrome measurement has been concluded for this round).
        \item Create a new array with timestamp $r=n+1$ which holds four `syndrome' plaquettes with `not remeasured' status. 
    \end{enumerate}
\end{enumerate}

The rules above implicitly use the fact that whenever the new check is measured, one of the plaquettes of the previous round is updated. This is guaranteed because subsequent checks are non-commuting.  For the same reason, one of the two plaquettes of the other color has to be randomized.  Thus, one can verify that for the rule 1(c), there will indeed be only one plaquette with `not remeasured' status at $r=n-T$. 
Additionally, we find by explicit verification that it is always possible to find a check that satisfies the requirement in 1(c). Similarly, if step 1(d) is reached, one can see that there will be at least two choices for checks. Thus, the algorithm cannot halt due to unsatisfiability of the requirements of the update rules.

Altogether, the protocol based on the $T$-PFA guarantees that for a single-qubit Pauli error, the first syndrome will be measured immediately after the error occurs, while the second one will be measured no later than $T$ rounds afterwards. This allows one to determine the spacetime location of the faulty edge and correct the error. In fact, additional information about errors is contained in checks because there are multiple ways to obtain the second syndrome from the measured checks (a plaquette can be formed in multiple ways by checks taken at various pairs of times; this applies e.g. to the last update of $P_b(Z)$ in Fig.~\ref{fig:PFA}).  Together with the correctability of single-qubit errors, this argues for the likelihood of fault-tolerance of either this protocol or at least some of its subclasses. It would be interesting to see if there exists an efficient decoder for dynamic tree codes generated by $T$-PFA and benchmark its performance. 

One might wonder if there exist nontrivial examples of such dynamic tree codes. In fact, random-flavor Floquet codes are the only solution for $T=3$, which, as we argued above, comprise a class of fault-tolerant dynamic codes. Another example are codes that follow the color sequence equivalent to $(r f_1 f_1 ) \left [(gf_2 f_2)(bf_3f_3) \right ]^s (gf_2 f_2)^{k=0,1}$ with $s + k \le T$, or symmetric versions thereof. An illustration of such  code is shown in Fig.~\ref{fig:PFA}.

\subsection{3D generalization}
It is a straightforward but cumbersome task to confirm that the random-flavor Floquet code in 3D, i.e. one that follows the $rgb  bgr$-like color sequence with consecutive checks anticommuting, corrects all single-qubit Pauli errors. The time signatures of the syndrome measurements are altered depending on the flavor sequence, however, which might affect the fault-tolerance properties of the code. Similarly, one can verify that a $T$-PFA approach can be, in principle, generalized to 3D.  Designing an efficient decoder for the 3D fracton Floquet code and 3D dynamic codes and analyzing their   fault tolerance properties would require a more involved analysis, so we leave this to a future work.

\section{Discussion}

In this paper, we presented several new dynamic codes in two and three dimensions that cannot be described under the subsystem code framework.  One immediately important direction would be to benchmark these codes and compare their performance to the honeycomb code with various error models.  
It should be possible to extend our analysis to finite Abelian groups, i.e. the case when the qubits are $\mathbb{Z}_N$ variables. Progress on this question has been made for the honeycomb code \cite{ZNpaper}and it would be useful to see if there are major qualitative differences for the CSS honeycomb code. 

Our protocols in 3D involve 3-body measurements, and it would be beneficial to find alternative constructions where measurements involve 2-body operators whilst preserving the error correcting properties of the fracton Floquet codes.  We found a preparation protocol for Haah's cubic code (shown in Appendix~\ref{app:Haahprep}) using two-body measurements, however constructing Floquet codes for type-II fractons would be very interesting.  Furthermore, fracton codes have been recently shown to have outstanding optimal thresholds for error correction \cite{https://doi.org/10.48550/arxiv.2112.05122}with the possibility of parallel error correction \cite{Brown_2020}. Therefore, it would  be interesting to rigorously benchmark the fracton Floquet code.

Another interesting question is the relation between the CSS honeycomb code and the e-m automorphism code from ref.~\cite{aasen}. Furthermore, one might ask whether there exists a unifying picture for dynamic tree codes and automorphism codes from the perspective of adiabatic paths of Hamiltonians, perhaps by utilizing the parent color code model. 

Finally, the dynamic tree codes proposed in this paper, especially the $T$-PFA generated codes, present an interesting way of constructing monitored random circuits using correlated randomness.   Understanding the robustness of this error correcting phase, and generalizing the code to a $T$-PFA construction that incorporates spatial nonuniformity would be valuable pursuits.  It would be curious to prove fault tolerance of these monitored random circuit codes by mapping to models of statistical mechanics.

\textit{Note Added:}While completing this manuscript, the authors became aware of an upcoming work where  two-dimensional CSS honeycomb codes are independently found from anyon condensation~[\onlinecite{brown_2022}], which provides a valuable framework for understanding Floquet codes, and also finds a numerical threshold for the code.

After completion of this manuscript, the authors learned of another forthcoming work~[\onlinecite{zhang_2022}] which introduces a fracton Floquet code with a codespace that grows with system size and a non-zero error threshold.

\section*{Acknowledgements}

We are grateful for useful discussions with Ike Chuang, Aram Harrow, Ali Lavasani, Michael Vasmer, Ethan Lake, John Preskill, Sagar Vijay, Dave Aasen and Dominic Williamson.  We are especially grateful for insightful discussions with Ben Brown.  N.T. is supported by the Walter Burke Institute for Theoretical Physics at Caltech.  S.B. is supported by the National Science Foundation Graduate Research Fellowship under Grant No. 1745302. This research was supported in part by the National Science Foundation under grant No. NSF PHY-1748958, the Heising-Simons Foundation, and the Simons Foundation (216179, LB).

\bibliography{ref}

\begin{thebibliography}{48}%
\makeatletter
\providecommand \@ifxundefined [1]{%
 \@ifx{#1\undefined}
}%
\providecommand \@ifnum [1]{%
 \ifnum #1\expandafter \@firstoftwo
 \else \expandafter \@secondoftwo
 \fi
}%
\providecommand \@ifx [1]{%
 \ifx #1\expandafter \@firstoftwo
 \else \expandafter \@secondoftwo
 \fi
}%
\providecommand \natexlab [1]{#1}%
\providecommand \enquote  [1]{``#1''}%
\providecommand \bibnamefont  [1]{#1}%
\providecommand \bibfnamefont [1]{#1}%
\providecommand \citenamefont [1]{#1}%
\providecommand \href@noop [0]{\@secondoftwo}%
\providecommand \href [0]{\begingroup \@sanitize@url \@href}%
\providecommand \@href[1]{\@@startlink{#1}\@@href}%
\providecommand \@@href[1]{\endgroup#1\@@endlink}%
\providecommand \@sanitize@url [0]{\catcode `\\12\catcode `\$12\catcode
  `\&12\catcode `\#12\catcode `\^12\catcode `\_12\catcode `\%12\relax}%
\providecommand \@@startlink[1]{}%
\providecommand \@@endlink[0]{}%
\providecommand \url  [0]{\begingroup\@sanitize@url \@url }%
\providecommand \@url [1]{\endgroup\@href {#1}{\urlprefix }}%
\providecommand \urlprefix  [0]{URL }%
\providecommand \Eprint [0]{\href }%
\providecommand \doibase [0]{https://doi.org/}%
\providecommand \selectlanguage [0]{\@gobble}%
\providecommand \bibinfo  [0]{\@secondoftwo}%
\providecommand \bibfield  [0]{\@secondoftwo}%
\providecommand \translation [1]{[#1]}%
\providecommand \BibitemOpen [0]{}%
\providecommand \bibitemStop [0]{}%
\providecommand \bibitemNoStop [0]{.\EOS\space}%
\providecommand \EOS [0]{\spacefactor3000\relax}%
\providecommand \BibitemShut  [1]{\csname bibitem#1\endcsname}%
\let\auto@bib@innerbib\@empty
%</preamble>
\bibitem [{\citenamefont {Poulin}(2005)}]{PhysRevLett.95.230504}%
  \BibitemOpen
  \bibfield  {author} {\bibinfo {author} {\bibfnamefont {D.}~\bibnamefont
  {Poulin}},\ }\bibfield  {title} {\bibinfo {title} {Stabilizer formalism for
  operator quantum error correction},\ }\href
  {https://doi.org/10.1103/PhysRevLett.95.230504} {\bibfield  {journal}
  {\bibinfo  {journal} {Phys. Rev. Lett.}\ }\textbf {\bibinfo {volume} {95}},\
  \bibinfo {pages} {230504} (\bibinfo {year} {2005})}\BibitemShut {NoStop}%
\bibitem [{\citenamefont {Bacon}(2006)}]{PhysRevA.73.012340}%
  \BibitemOpen
  \bibfield  {author} {\bibinfo {author} {\bibfnamefont {D.}~\bibnamefont
  {Bacon}},\ }\bibfield  {title} {\bibinfo {title} {Operator quantum
  error-correcting subsystems for self-correcting quantum memories},\ }\href
  {https://doi.org/10.1103/PhysRevA.73.012340} {\bibfield  {journal} {\bibinfo
  {journal} {Phys. Rev. A}\ }\textbf {\bibinfo {volume} {73}},\ \bibinfo
  {pages} {012340} (\bibinfo {year} {2006})}\BibitemShut {NoStop}%
\bibitem [{\citenamefont {Bombin}(2010)}]{PhysRevA.81.032301}%
  \BibitemOpen
  \bibfield  {author} {\bibinfo {author} {\bibfnamefont {H.}~\bibnamefont
  {Bombin}},\ }\bibfield  {title} {\bibinfo {title} {Topological subsystem
  codes},\ }\href {https://doi.org/10.1103/PhysRevA.81.032301} {\bibfield
  {journal} {\bibinfo  {journal} {Phys. Rev. A}\ }\textbf {\bibinfo {volume}
  {81}},\ \bibinfo {pages} {032301} (\bibinfo {year} {2010})}\BibitemShut
  {NoStop}%
\bibitem [{\citenamefont {Gottesman}(1998)}]{Gottesman_1998}%
  \BibitemOpen
  \bibfield  {author} {\bibinfo {author} {\bibfnamefont {D.}~\bibnamefont
  {Gottesman}},\ }\bibfield  {title} {\bibinfo {title} {Theory of
  fault-tolerant quantum computation},\ }\href
  {https://doi.org/10.1103/physreva.57.127} {\bibfield  {journal} {\bibinfo
  {journal} {Physical Review A}\ }\textbf {\bibinfo {volume} {57}},\ \bibinfo
  {pages} {127} (\bibinfo {year} {1998})}\BibitemShut {NoStop}%
\bibitem [{\citenamefont {Kubica}\ and\ \citenamefont
  {Beverland}(2015)}]{PhysRevA.91.032330}%
  \BibitemOpen
  \bibfield  {author} {\bibinfo {author} {\bibfnamefont {A.}~\bibnamefont
  {Kubica}}\ and\ \bibinfo {author} {\bibfnamefont {M.~E.}\ \bibnamefont
  {Beverland}},\ }\bibfield  {title} {\bibinfo {title} {Universal transversal
  gates with color codes: A simplified approach},\ }\href
  {https://doi.org/10.1103/PhysRevA.91.032330} {\bibfield  {journal} {\bibinfo
  {journal} {Phys. Rev. A}\ }\textbf {\bibinfo {volume} {91}},\ \bibinfo
  {pages} {032330} (\bibinfo {year} {2015})}\BibitemShut {NoStop}%
\bibitem [{\citenamefont {Bomb{\'\i}n}(2015)}]{bombin2015gauge}%
  \BibitemOpen
  \bibfield  {author} {\bibinfo {author} {\bibfnamefont {H.}~\bibnamefont
  {Bomb{\'\i}n}},\ }\bibfield  {title} {\bibinfo {title} {Gauge color codes:
  optimal transversal gates and gauge fixing in topological stabilizer codes},\
  }\href {https://iopscience.iop.org/article/10.1088/1367-2630/17/8/083002}
  {\bibfield  {journal} {\bibinfo  {journal} {New Journal of Physics}\ }\textbf
  {\bibinfo {volume} {17}},\ \bibinfo {pages} {083002} (\bibinfo {year}
  {2015})}\BibitemShut {NoStop}%
\bibitem [{\citenamefont {Anderson}\ \emph {et~al.}(2014)\citenamefont
  {Anderson}, \citenamefont {Duclos-Cianci},\ and\ \citenamefont
  {Poulin}}]{PhysRevLett.113.080501}%
  \BibitemOpen
  \bibfield  {author} {\bibinfo {author} {\bibfnamefont {J.~T.}\ \bibnamefont
  {Anderson}}, \bibinfo {author} {\bibfnamefont {G.}~\bibnamefont
  {Duclos-Cianci}},\ and\ \bibinfo {author} {\bibfnamefont {D.}~\bibnamefont
  {Poulin}},\ }\bibfield  {title} {\bibinfo {title} {Fault-tolerant conversion
  between the steane and reed-muller quantum codes},\ }\href
  {https://doi.org/10.1103/PhysRevLett.113.080501} {\bibfield  {journal}
  {\bibinfo  {journal} {Phys. Rev. Lett.}\ }\textbf {\bibinfo {volume} {113}},\
  \bibinfo {pages} {080501} (\bibinfo {year} {2014})}\BibitemShut {NoStop}%
\bibitem [{\citenamefont {Paetznick}\ and\ \citenamefont
  {Reichardt}(2013)}]{PhysRevLett.111.090505}%
  \BibitemOpen
  \bibfield  {author} {\bibinfo {author} {\bibfnamefont {A.}~\bibnamefont
  {Paetznick}}\ and\ \bibinfo {author} {\bibfnamefont {B.~W.}\ \bibnamefont
  {Reichardt}},\ }\bibfield  {title} {\bibinfo {title} {Universal
  fault-tolerant quantum computation with only transversal gates and error
  correction},\ }\href {https://doi.org/10.1103/PhysRevLett.111.090505}
  {\bibfield  {journal} {\bibinfo  {journal} {Phys. Rev. Lett.}\ }\textbf
  {\bibinfo {volume} {111}},\ \bibinfo {pages} {090505} (\bibinfo {year}
  {2013})}\BibitemShut {NoStop}%
\bibitem [{\citenamefont {Eastin}\ and\ \citenamefont
  {Knill}(2009)}]{PhysRevLett.102.110502}%
  \BibitemOpen
  \bibfield  {author} {\bibinfo {author} {\bibfnamefont {B.}~\bibnamefont
  {Eastin}}\ and\ \bibinfo {author} {\bibfnamefont {E.}~\bibnamefont {Knill}},\
  }\bibfield  {title} {\bibinfo {title} {Restrictions on transversal encoded
  quantum gate sets},\ }\href {https://doi.org/10.1103/PhysRevLett.102.110502}
  {\bibfield  {journal} {\bibinfo  {journal} {Phys. Rev. Lett.}\ }\textbf
  {\bibinfo {volume} {102}},\ \bibinfo {pages} {110502} (\bibinfo {year}
  {2009})}\BibitemShut {NoStop}%
\bibitem [{\citenamefont {Webster}\ \emph {et~al.}(2022)\citenamefont
  {Webster}, \citenamefont {Vasmer}, \citenamefont {Scruby},\ and\
  \citenamefont {Bartlett}}]{PhysRevResearch.4.013092}%
  \BibitemOpen
  \bibfield  {author} {\bibinfo {author} {\bibfnamefont {P.}~\bibnamefont
  {Webster}}, \bibinfo {author} {\bibfnamefont {M.}~\bibnamefont {Vasmer}},
  \bibinfo {author} {\bibfnamefont {T.~R.}\ \bibnamefont {Scruby}},\ and\
  \bibinfo {author} {\bibfnamefont {S.~D.}\ \bibnamefont {Bartlett}},\
  }\bibfield  {title} {\bibinfo {title} {Universal fault-tolerant quantum
  computing with stabilizer codes},\ }\href
  {https://doi.org/10.1103/PhysRevResearch.4.013092} {\bibfield  {journal}
  {\bibinfo  {journal} {Phys. Rev. Research}\ }\textbf {\bibinfo {volume}
  {4}},\ \bibinfo {pages} {013092} (\bibinfo {year} {2022})}\BibitemShut
  {NoStop}%
\bibitem [{\citenamefont {Bomb{\'\i}n}\ and\ \citenamefont
  {Martin-Delgado}(2009)}]{bombin2009quantum}%
  \BibitemOpen
  \bibfield  {author} {\bibinfo {author} {\bibfnamefont {H.}~\bibnamefont
  {Bomb{\'\i}n}}\ and\ \bibinfo {author} {\bibfnamefont {M.~A.}\ \bibnamefont
  {Martin-Delgado}},\ }\bibfield  {title} {\bibinfo {title} {Quantum
  measurements and gates by code deformation},\ }\href
  {https://iopscience.iop.org/article/10.1088/1751-8113/42/9/095302} {\bibfield
   {journal} {\bibinfo  {journal} {Journal of Physics A: Mathematical and
  Theoretical}\ }\textbf {\bibinfo {volume} {42}},\ \bibinfo {pages} {095302}
  (\bibinfo {year} {2009})}\BibitemShut {NoStop}%
\bibitem [{\citenamefont {Horsman}\ \emph {et~al.}(2012)\citenamefont
  {Horsman}, \citenamefont {Fowler}, \citenamefont {Devitt},\ and\
  \citenamefont {Van~Meter}}]{horsman2012surface}%
  \BibitemOpen
  \bibfield  {author} {\bibinfo {author} {\bibfnamefont {C.}~\bibnamefont
  {Horsman}}, \bibinfo {author} {\bibfnamefont {A.~G.}\ \bibnamefont {Fowler}},
  \bibinfo {author} {\bibfnamefont {S.}~\bibnamefont {Devitt}},\ and\ \bibinfo
  {author} {\bibfnamefont {R.}~\bibnamefont {Van~Meter}},\ }\bibfield  {title}
  {\bibinfo {title} {Surface code quantum computing by lattice surgery},\
  }\href {https://iopscience.iop.org/article/10.1088/1367-2630/14/12/123011}
  {\bibfield  {journal} {\bibinfo  {journal} {New Journal of Physics}\ }\textbf
  {\bibinfo {volume} {14}},\ \bibinfo {pages} {123011} (\bibinfo {year}
  {2012})}\BibitemShut {NoStop}%
\bibitem [{\citenamefont {Vuillot}\ \emph {et~al.}(2019)\citenamefont
  {Vuillot}, \citenamefont {Lao}, \citenamefont {Criger}, \citenamefont
  {Almud{\'e}ver}, \citenamefont {Bertels},\ and\ \citenamefont
  {Terhal}}]{vuillot2019code}%
  \BibitemOpen
  \bibfield  {author} {\bibinfo {author} {\bibfnamefont {C.}~\bibnamefont
  {Vuillot}}, \bibinfo {author} {\bibfnamefont {L.}~\bibnamefont {Lao}},
  \bibinfo {author} {\bibfnamefont {B.}~\bibnamefont {Criger}}, \bibinfo
  {author} {\bibfnamefont {C.~G.}\ \bibnamefont {Almud{\'e}ver}}, \bibinfo
  {author} {\bibfnamefont {K.}~\bibnamefont {Bertels}},\ and\ \bibinfo {author}
  {\bibfnamefont {B.~M.}\ \bibnamefont {Terhal}},\ }\bibfield  {title}
  {\bibinfo {title} {Code deformation and lattice surgery are gauge fixing},\
  }\href {https://iopscience.iop.org/article/10.1088/1367-2630/ab0199/meta}
  {\bibfield  {journal} {\bibinfo  {journal} {New Journal of Physics}\ }\textbf
  {\bibinfo {volume} {21}},\ \bibinfo {pages} {033028} (\bibinfo {year}
  {2019})}\BibitemShut {NoStop}%
\bibitem [{\citenamefont {Hastings}\ and\ \citenamefont
  {Haah}(2021)}]{hastings2021dynamically}%
  \BibitemOpen
  \bibfield  {author} {\bibinfo {author} {\bibfnamefont {M.~B.}\ \bibnamefont
  {Hastings}}\ and\ \bibinfo {author} {\bibfnamefont {J.}~\bibnamefont
  {Haah}},\ }\bibfield  {title} {\bibinfo {title} {Dynamically {G}enerated
  {L}ogical {Q}ubits},\ }\href {https://doi.org/10.22331/q-2021-10-19-564}
  {\bibfield  {journal} {\bibinfo  {journal} {{Quantum}}\ }\textbf {\bibinfo
  {volume} {5}},\ \bibinfo {pages} {564} (\bibinfo {year} {2021})}\BibitemShut
  {NoStop}%
\bibitem [{\citenamefont {Haah}\ and\ \citenamefont
  {Hastings}(2022)}]{Haah_2022}%
  \BibitemOpen
  \bibfield  {author} {\bibinfo {author} {\bibfnamefont {J.}~\bibnamefont
  {Haah}}\ and\ \bibinfo {author} {\bibfnamefont {M.~B.}\ \bibnamefont
  {Hastings}},\ }\bibfield  {title} {\bibinfo {title} {Boundaries for the
  honeycomb code},\ }\href {https://doi.org/10.22331/q-2022-04-21-693}
  {\bibfield  {journal} {\bibinfo  {journal} {Quantum}\ }\textbf {\bibinfo
  {volume} {6}},\ \bibinfo {pages} {693} (\bibinfo {year} {2022})}\BibitemShut
  {NoStop}%
\bibitem [{\citenamefont {Kitaev}(2006)}]{kitaev2006anyons}%
  \BibitemOpen
  \bibfield  {author} {\bibinfo {author} {\bibfnamefont {A.}~\bibnamefont
  {Kitaev}},\ }\bibfield  {title} {\bibinfo {title} {Anyons in an exactly
  solved model and beyond},\ }\href
  {https://www.sciencedirect.com/science/article/pii/S0003491605002381?via%3Dihub}
  {\bibfield  {journal} {\bibinfo  {journal} {Annals of Physics}\ }\textbf
  {\bibinfo {volume} {321}},\ \bibinfo {pages} {2} (\bibinfo {year}
  {2006})}\BibitemShut {NoStop}%
\bibitem [{\citenamefont {Suchara}\ \emph {et~al.}(2011)\citenamefont
  {Suchara}, \citenamefont {Bravyi},\ and\ \citenamefont
  {Terhal}}]{suchara2011constructions}%
  \BibitemOpen
  \bibfield  {author} {\bibinfo {author} {\bibfnamefont {M.}~\bibnamefont
  {Suchara}}, \bibinfo {author} {\bibfnamefont {S.}~\bibnamefont {Bravyi}},\
  and\ \bibinfo {author} {\bibfnamefont {B.}~\bibnamefont {Terhal}},\
  }\bibfield  {title} {\bibinfo {title} {Constructions and noise threshold of
  topological subsystem codes},\ }\href
  {https://iopscience.iop.org/article/10.1088/1751-8113/44/15/155301/meta}
  {\bibfield  {journal} {\bibinfo  {journal} {Journal of Physics A:
  Mathematical and Theoretical}\ }\textbf {\bibinfo {volume} {44}},\ \bibinfo
  {pages} {155301} (\bibinfo {year} {2011})}\BibitemShut {NoStop}%
\bibitem [{\citenamefont {Kitaev}(2003)}]{kitaev2003fault}%
  \BibitemOpen
  \bibfield  {author} {\bibinfo {author} {\bibfnamefont {A.~Y.}\ \bibnamefont
  {Kitaev}},\ }\bibfield  {title} {\bibinfo {title} {Fault-tolerant quantum
  computation by anyons},\ }\href
  {https://www.sciencedirect.com/science/article/pii/S0003491602000180}
  {\bibfield  {journal} {\bibinfo  {journal} {Annals of Physics}\ }\textbf
  {\bibinfo {volume} {303}},\ \bibinfo {pages} {2} (\bibinfo {year}
  {2003})}\BibitemShut {NoStop}%
\bibitem [{\citenamefont {Gidney}\ \emph {et~al.}(2021)\citenamefont {Gidney},
  \citenamefont {Newman}, \citenamefont {Fowler},\ and\ \citenamefont
  {Broughton}}]{Gidney2021faulttolerant}%
  \BibitemOpen
  \bibfield  {author} {\bibinfo {author} {\bibfnamefont {C.}~\bibnamefont
  {Gidney}}, \bibinfo {author} {\bibfnamefont {M.}~\bibnamefont {Newman}},
  \bibinfo {author} {\bibfnamefont {A.}~\bibnamefont {Fowler}},\ and\ \bibinfo
  {author} {\bibfnamefont {M.}~\bibnamefont {Broughton}},\ }\bibfield  {title}
  {\bibinfo {title} {A {F}ault-{T}olerant {H}oneycomb {M}emory},\ }\href
  {https://doi.org/10.22331/q-2021-12-20-605} {\bibfield  {journal} {\bibinfo
  {journal} {{Quantum}}\ }\textbf {\bibinfo {volume} {5}},\ \bibinfo {pages}
  {605} (\bibinfo {year} {2021})}\BibitemShut {NoStop}%
\bibitem [{\citenamefont {Gidney}\ \emph {et~al.}(2022)\citenamefont {Gidney},
  \citenamefont {Newman},\ and\ \citenamefont
  {McEwen}}]{gidney2022benchmarking}%
  \BibitemOpen
  \bibfield  {author} {\bibinfo {author} {\bibfnamefont {C.}~\bibnamefont
  {Gidney}}, \bibinfo {author} {\bibfnamefont {M.}~\bibnamefont {Newman}},\
  and\ \bibinfo {author} {\bibfnamefont {M.}~\bibnamefont {McEwen}},\
  }\bibfield  {title} {\bibinfo {title} {Benchmarking the {P}lanar {H}oneycomb
  {C}ode},\ }\href {https://doi.org/10.22331/q-2022-09-21-813} {\bibfield
  {journal} {\bibinfo  {journal} {{Quantum}}\ }\textbf {\bibinfo {volume}
  {6}},\ \bibinfo {pages} {813} (\bibinfo {year} {2022})}\BibitemShut {NoStop}%
\bibitem [{\citenamefont {Aasen}\ \emph {et~al.}(2022)\citenamefont {Aasen},
  \citenamefont {Wang},\ and\ \citenamefont {Hastings}}]{aasen}%
  \BibitemOpen
  \bibfield  {author} {\bibinfo {author} {\bibfnamefont {D.}~\bibnamefont
  {Aasen}}, \bibinfo {author} {\bibfnamefont {Z.}~\bibnamefont {Wang}},\ and\
  \bibinfo {author} {\bibfnamefont {M.~B.}\ \bibnamefont {Hastings}},\
  }\bibfield  {title} {\bibinfo {title} {Adiabatic paths of hamiltonians,
  symmetries of topological order, and automorphism codes},\ }\href
  {https://doi.org/10.1103/PhysRevB.106.085122} {\bibfield  {journal} {\bibinfo
   {journal} {Phys. Rev. B}\ }\textbf {\bibinfo {volume} {106}},\ \bibinfo
  {pages} {085122} (\bibinfo {year} {2022})}\BibitemShut {NoStop}%
\bibitem [{\citenamefont {Vijay}\ \emph
  {et~al.}(2016{\natexlab{a}})\citenamefont {Vijay}, \citenamefont {Haah},\
  and\ \citenamefont {Fu}}]{PhysRevB.94.235157}%
  \BibitemOpen
  \bibfield  {author} {\bibinfo {author} {\bibfnamefont {S.}~\bibnamefont
  {Vijay}}, \bibinfo {author} {\bibfnamefont {J.}~\bibnamefont {Haah}},\ and\
  \bibinfo {author} {\bibfnamefont {L.}~\bibnamefont {Fu}},\ }\bibfield
  {title} {\bibinfo {title} {Fracton topological order, generalized lattice
  gauge theory, and duality},\ }\href
  {https://doi.org/10.1103/PhysRevB.94.235157} {\bibfield  {journal} {\bibinfo
  {journal} {Phys. Rev. B}\ }\textbf {\bibinfo {volume} {94}},\ \bibinfo
  {pages} {235157} (\bibinfo {year} {2016}{\natexlab{a}})}\BibitemShut
  {NoStop}%
\bibitem [{\citenamefont {Balasubramanian}\ \emph
  {et~al.}(2022{\natexlab{a}})\citenamefont {Balasubramanian}, \citenamefont
  {Galitski},\ and\ \citenamefont {Vishwanath}}]{Balasubramanian}%
  \BibitemOpen
  \bibfield  {author} {\bibinfo {author} {\bibfnamefont {S.}~\bibnamefont
  {Balasubramanian}}, \bibinfo {author} {\bibfnamefont {V.}~\bibnamefont
  {Galitski}},\ and\ \bibinfo {author} {\bibfnamefont {A.}~\bibnamefont
  {Vishwanath}},\ }\bibfield  {title} {\bibinfo {title} {Classical vertex model
  dualities in a family of two-dimensional frustrated quantum
  antiferromagnets},\ }\href {https://doi.org/10.1103/PhysRevB.106.195127}
  {\bibfield  {journal} {\bibinfo  {journal} {Phys. Rev. B}\ }\textbf {\bibinfo
  {volume} {106}},\ \bibinfo {pages} {195127} (\bibinfo {year}
  {2022}{\natexlab{a}})}\BibitemShut {NoStop}%
\bibitem [{\citenamefont {Balasubramanian}\ \emph
  {et~al.}(2022{\natexlab{b}})\citenamefont {Balasubramanian}, \citenamefont
  {Bulmash}, \citenamefont {Galitski},\ and\ \citenamefont
  {Vishwanath}}]{Balasubramanian2}%
  \BibitemOpen
  \bibfield  {author} {\bibinfo {author} {\bibfnamefont {S.}~\bibnamefont
  {Balasubramanian}}, \bibinfo {author} {\bibfnamefont {D.}~\bibnamefont
  {Bulmash}}, \bibinfo {author} {\bibfnamefont {V.}~\bibnamefont {Galitski}},\
  and\ \bibinfo {author} {\bibfnamefont {A.}~\bibnamefont {Vishwanath}},\
  }\bibfield  {title} {\bibinfo {title} {Exact wavefunction dualities and phase
  diagrams of 3d quantum vertex models},\ }\href
  {https://arxiv.org/abs/2201.08856} {\bibfield  {journal} {\bibinfo  {journal}
  {arXiv preprint arXiv:2201.08856}\ } (\bibinfo {year}
  {2022}{\natexlab{b}})}\BibitemShut {NoStop}%
\bibitem [{\citenamefont {Vuillot}(2021)}]{vuillot}%
  \BibitemOpen
  \bibfield  {author} {\bibinfo {author} {\bibfnamefont {C.}~\bibnamefont
  {Vuillot}},\ }\href {https://doi.org/10.48550/arxiv.2110.05348} {\bibinfo
  {title} {Planar floquet codes}} (\bibinfo {year} {2021})\BibitemShut
  {NoStop}%
\bibitem [{\citenamefont {Wen}(2003)}]{PhysRevLett.90.016803}%
  \BibitemOpen
  \bibfield  {author} {\bibinfo {author} {\bibfnamefont {X.-G.}\ \bibnamefont
  {Wen}},\ }\bibfield  {title} {\bibinfo {title} {Quantum orders in an exact
  soluble model},\ }\href {https://doi.org/10.1103/PhysRevLett.90.016803}
  {\bibfield  {journal} {\bibinfo  {journal} {Phys. Rev. Lett.}\ }\textbf
  {\bibinfo {volume} {90}},\ \bibinfo {pages} {016803} (\bibinfo {year}
  {2003})}\BibitemShut {NoStop}%
\bibitem [{\citenamefont {Kesselring}\ \emph {et~al.}(2022)\citenamefont
  {Kesselring}, \citenamefont {de~la Fuente}, \citenamefont {Thomsen},
  \citenamefont {Eisert}, \citenamefont {Bartlett},\ and\ \citenamefont
  {Brown}}]{brown_2022}%
  \BibitemOpen
  \bibfield  {author} {\bibinfo {author} {\bibfnamefont {M.~S.}\ \bibnamefont
  {Kesselring}}, \bibinfo {author} {\bibfnamefont {J.~C.~M.}\ \bibnamefont
  {de~la Fuente}}, \bibinfo {author} {\bibfnamefont {F.}~\bibnamefont
  {Thomsen}}, \bibinfo {author} {\bibfnamefont {J.}~\bibnamefont {Eisert}},
  \bibinfo {author} {\bibfnamefont {S.~D.}\ \bibnamefont {Bartlett}},\ and\
  \bibinfo {author} {\bibfnamefont {B.~J.}\ \bibnamefont {Brown}},\ }\href
  {https://doi.org/10.48550/ARXIV.2212.00042} {\bibinfo {title} {Anyon
  condensation and the color code}} (\bibinfo {year} {2022})\BibitemShut
  {NoStop}%
\bibitem [{\citenamefont {Dennis}\ \emph {et~al.}(2002)\citenamefont {Dennis},
  \citenamefont {Kitaev}, \citenamefont {Landahl},\ and\ \citenamefont
  {Preskill}}]{dennis2002topological}%
  \BibitemOpen
  \bibfield  {author} {\bibinfo {author} {\bibfnamefont {E.}~\bibnamefont
  {Dennis}}, \bibinfo {author} {\bibfnamefont {A.}~\bibnamefont {Kitaev}},
  \bibinfo {author} {\bibfnamefont {A.}~\bibnamefont {Landahl}},\ and\ \bibinfo
  {author} {\bibfnamefont {J.}~\bibnamefont {Preskill}},\ }\bibfield  {title}
  {\bibinfo {title} {Topological quantum memory},\ }\href
  {https://aip.scitation.org/doi/10.1063/1.1499754} {\bibfield  {journal}
  {\bibinfo  {journal} {Journal of Mathematical Physics}\ }\textbf {\bibinfo
  {volume} {43}},\ \bibinfo {pages} {4452} (\bibinfo {year}
  {2002})}\BibitemShut {NoStop}%
\bibitem [{\citenamefont {Vijay}\ \emph
  {et~al.}(2016{\natexlab{b}})\citenamefont {Vijay}, \citenamefont {Haah},\
  and\ \citenamefont {Fu}}]{VijayHaahFu16}%
  \BibitemOpen
  \bibfield  {author} {\bibinfo {author} {\bibfnamefont {S.}~\bibnamefont
  {Vijay}}, \bibinfo {author} {\bibfnamefont {J.}~\bibnamefont {Haah}},\ and\
  \bibinfo {author} {\bibfnamefont {L.}~\bibnamefont {Fu}},\ }\bibfield
  {title} {\bibinfo {title} {Fracton topological order, generalized lattice
  gauge theory, and duality},\ }\href
  {https://doi.org/10.1103/PhysRevB.94.235157} {\bibfield  {journal} {\bibinfo
  {journal} {Phys. Rev. B}\ }\textbf {\bibinfo {volume} {94}},\ \bibinfo
  {pages} {235157} (\bibinfo {year} {2016}{\natexlab{b}})}\BibitemShut
  {NoStop}%
\bibitem [{\citenamefont {Shirley}\ \emph {et~al.}(2019)\citenamefont
  {Shirley}, \citenamefont {Slagle},\ and\ \citenamefont
  {Chen}}]{PhysRevB.99.115123}%
  \BibitemOpen
  \bibfield  {author} {\bibinfo {author} {\bibfnamefont {W.}~\bibnamefont
  {Shirley}}, \bibinfo {author} {\bibfnamefont {K.}~\bibnamefont {Slagle}},\
  and\ \bibinfo {author} {\bibfnamefont {X.}~\bibnamefont {Chen}},\ }\bibfield
  {title} {\bibinfo {title} {Foliated fracton order in the checkerboard
  model},\ }\href {https://doi.org/10.1103/PhysRevB.99.115123} {\bibfield
  {journal} {\bibinfo  {journal} {Phys. Rev. B}\ }\textbf {\bibinfo {volume}
  {99}},\ \bibinfo {pages} {115123} (\bibinfo {year} {2019})}\BibitemShut
  {NoStop}%
\bibitem [{\citenamefont {Kubica}\ and\ \citenamefont
  {Vasmer}(2022)}]{kubica2022single}%
  \BibitemOpen
  \bibfield  {author} {\bibinfo {author} {\bibfnamefont {A.}~\bibnamefont
  {Kubica}}\ and\ \bibinfo {author} {\bibfnamefont {M.}~\bibnamefont
  {Vasmer}},\ }\bibfield  {title} {\bibinfo {title} {Single-shot quantum error
  correction with the three-dimensional subsystem toric code},\ }\href@noop {}
  {\bibfield  {journal} {\bibinfo  {journal} {Nature communications}\ }\textbf
  {\bibinfo {volume} {13}},\ \bibinfo {pages} {1} (\bibinfo {year}
  {2022})}\BibitemShut {NoStop}%
\bibitem [{\citenamefont {Song}\ \emph {et~al.}(2021)\citenamefont {Song},
  \citenamefont {Schönmeier-Kromer}, \citenamefont {Liu}, \citenamefont
  {Viyuela}, \citenamefont {Pollet},\ and\ \citenamefont
  {Martin-Delgado}}]{https://doi.org/10.48550/arxiv.2112.05122}%
  \BibitemOpen
  \bibfield  {author} {\bibinfo {author} {\bibfnamefont {H.}~\bibnamefont
  {Song}}, \bibinfo {author} {\bibfnamefont {J.}~\bibnamefont
  {Schönmeier-Kromer}}, \bibinfo {author} {\bibfnamefont {K.}~\bibnamefont
  {Liu}}, \bibinfo {author} {\bibfnamefont {O.}~\bibnamefont {Viyuela}},
  \bibinfo {author} {\bibfnamefont {L.}~\bibnamefont {Pollet}},\ and\ \bibinfo
  {author} {\bibfnamefont {M.~A.}\ \bibnamefont {Martin-Delgado}},\ }\bibfield
  {title} {\bibinfo {title} {Optimal thresholds for fracton codes and random
  spin models with subsystem symmetry}\ }\href
  {https://doi.org/10.48550/arxiv.2112.05122} {10.48550/arxiv.2112.05122}
  (\bibinfo {year} {2021})\BibitemShut {NoStop}%
\bibitem [{\citenamefont {Skinner}\ \emph {et~al.}(2019)\citenamefont
  {Skinner}, \citenamefont {Ruhman},\ and\ \citenamefont
  {Nahum}}]{PhysRevX.9.031009}%
  \BibitemOpen
  \bibfield  {author} {\bibinfo {author} {\bibfnamefont {B.}~\bibnamefont
  {Skinner}}, \bibinfo {author} {\bibfnamefont {J.}~\bibnamefont {Ruhman}},\
  and\ \bibinfo {author} {\bibfnamefont {A.}~\bibnamefont {Nahum}},\ }\bibfield
   {title} {\bibinfo {title} {Measurement-induced phase transitions in the
  dynamics of entanglement},\ }\href
  {https://doi.org/10.1103/PhysRevX.9.031009} {\bibfield  {journal} {\bibinfo
  {journal} {Phys. Rev. X}\ }\textbf {\bibinfo {volume} {9}},\ \bibinfo {pages}
  {031009} (\bibinfo {year} {2019})}\BibitemShut {NoStop}%
\bibitem [{\citenamefont {Gullans}\ and\ \citenamefont
  {Huse}(2020)}]{PhysRevX.10.041020}%
  \BibitemOpen
  \bibfield  {author} {\bibinfo {author} {\bibfnamefont {M.~J.}\ \bibnamefont
  {Gullans}}\ and\ \bibinfo {author} {\bibfnamefont {D.~A.}\ \bibnamefont
  {Huse}},\ }\bibfield  {title} {\bibinfo {title} {Dynamical purification phase
  transition induced by quantum measurements},\ }\href
  {https://doi.org/10.1103/PhysRevX.10.041020} {\bibfield  {journal} {\bibinfo
  {journal} {Phys. Rev. X}\ }\textbf {\bibinfo {volume} {10}},\ \bibinfo
  {pages} {041020} (\bibinfo {year} {2020})}\BibitemShut {NoStop}%
\bibitem [{\citenamefont {Lavasani}\ \emph {et~al.}(2021)\citenamefont
  {Lavasani}, \citenamefont {Alavirad},\ and\ \citenamefont
  {Barkeshli}}]{PhysRevLett.127.235701}%
  \BibitemOpen
  \bibfield  {author} {\bibinfo {author} {\bibfnamefont {A.}~\bibnamefont
  {Lavasani}}, \bibinfo {author} {\bibfnamefont {Y.}~\bibnamefont {Alavirad}},\
  and\ \bibinfo {author} {\bibfnamefont {M.}~\bibnamefont {Barkeshli}},\
  }\bibfield  {title} {\bibinfo {title} {Topological order and criticality in
  $(2+1)\mathrm{D}$ monitored random quantum circuits},\ }\href
  {https://doi.org/10.1103/PhysRevLett.127.235701} {\bibfield  {journal}
  {\bibinfo  {journal} {Phys. Rev. Lett.}\ }\textbf {\bibinfo {volume} {127}},\
  \bibinfo {pages} {235701} (\bibinfo {year} {2021})}\BibitemShut {NoStop}%
\bibitem [{\citenamefont {Fan}\ \emph {et~al.}(2021)\citenamefont {Fan},
  \citenamefont {Vijay}, \citenamefont {Vishwanath},\ and\ \citenamefont
  {You}}]{PhysRevB.103.174309}%
  \BibitemOpen
  \bibfield  {author} {\bibinfo {author} {\bibfnamefont {R.}~\bibnamefont
  {Fan}}, \bibinfo {author} {\bibfnamefont {S.}~\bibnamefont {Vijay}}, \bibinfo
  {author} {\bibfnamefont {A.}~\bibnamefont {Vishwanath}},\ and\ \bibinfo
  {author} {\bibfnamefont {Y.-Z.}\ \bibnamefont {You}},\ }\bibfield  {title}
  {\bibinfo {title} {Self-organized error correction in random unitary circuits
  with measurement},\ }\href {https://doi.org/10.1103/PhysRevB.103.174309}
  {\bibfield  {journal} {\bibinfo  {journal} {Phys. Rev. B}\ }\textbf {\bibinfo
  {volume} {103}},\ \bibinfo {pages} {174309} (\bibinfo {year}
  {2021})}\BibitemShut {NoStop}%
\bibitem [{\citenamefont {Li}\ \emph {et~al.}(2018)\citenamefont {Li},
  \citenamefont {Chen},\ and\ \citenamefont {Fisher}}]{PhysRevB.98.205136}%
  \BibitemOpen
  \bibfield  {author} {\bibinfo {author} {\bibfnamefont {Y.}~\bibnamefont
  {Li}}, \bibinfo {author} {\bibfnamefont {X.}~\bibnamefont {Chen}},\ and\
  \bibinfo {author} {\bibfnamefont {M.~P.~A.}\ \bibnamefont {Fisher}},\
  }\bibfield  {title} {\bibinfo {title} {Quantum zeno effect and the many-body
  entanglement transition},\ }\href
  {https://doi.org/10.1103/PhysRevB.98.205136} {\bibfield  {journal} {\bibinfo
  {journal} {Phys. Rev. B}\ }\textbf {\bibinfo {volume} {98}},\ \bibinfo
  {pages} {205136} (\bibinfo {year} {2018})}\BibitemShut {NoStop}%
\bibitem [{\citenamefont {Choi}\ \emph {et~al.}(2020)\citenamefont {Choi},
  \citenamefont {Bao}, \citenamefont {Qi},\ and\ \citenamefont
  {Altman}}]{PhysRevLett.125.030505}%
  \BibitemOpen
  \bibfield  {author} {\bibinfo {author} {\bibfnamefont {S.}~\bibnamefont
  {Choi}}, \bibinfo {author} {\bibfnamefont {Y.}~\bibnamefont {Bao}}, \bibinfo
  {author} {\bibfnamefont {X.-L.}\ \bibnamefont {Qi}},\ and\ \bibinfo {author}
  {\bibfnamefont {E.}~\bibnamefont {Altman}},\ }\bibfield  {title} {\bibinfo
  {title} {Quantum error correction in scrambling dynamics and
  measurement-induced phase transition},\ }\href
  {https://doi.org/10.1103/PhysRevLett.125.030505} {\bibfield  {journal}
  {\bibinfo  {journal} {Phys. Rev. Lett.}\ }\textbf {\bibinfo {volume} {125}},\
  \bibinfo {pages} {030505} (\bibinfo {year} {2020})}\BibitemShut {NoStop}%
\bibitem [{\citenamefont {Li}\ and\ \citenamefont
  {Fisher}(2021)}]{PhysRevB.103.104306}%
  \BibitemOpen
  \bibfield  {author} {\bibinfo {author} {\bibfnamefont {Y.}~\bibnamefont
  {Li}}\ and\ \bibinfo {author} {\bibfnamefont {M.~P.~A.}\ \bibnamefont
  {Fisher}},\ }\bibfield  {title} {\bibinfo {title} {Statistical mechanics of
  quantum error correcting codes},\ }\href
  {https://doi.org/10.1103/PhysRevB.103.104306} {\bibfield  {journal} {\bibinfo
   {journal} {Phys. Rev. B}\ }\textbf {\bibinfo {volume} {103}},\ \bibinfo
  {pages} {104306} (\bibinfo {year} {2021})}\BibitemShut {NoStop}%
\bibitem [{\citenamefont {Li}\ \emph {et~al.}(2023)\citenamefont {Li},
  \citenamefont {Vijay},\ and\ \citenamefont
  {Fisher}}]{https://doi.org/10.48550/arxiv.2105.13352}%
  \BibitemOpen
  \bibfield  {author} {\bibinfo {author} {\bibfnamefont {Y.}~\bibnamefont
  {Li}}, \bibinfo {author} {\bibfnamefont {S.}~\bibnamefont {Vijay}},\ and\
  \bibinfo {author} {\bibfnamefont {M.~P.}\ \bibnamefont {Fisher}},\ }\bibfield
   {title} {\bibinfo {title} {Entanglement domain walls in monitored quantum
  circuits and the directed polymer in a random environment},\ }\href
  {https://doi.org/10.1103/PRXQuantum.4.010331} {\bibfield  {journal} {\bibinfo
   {journal} {PRX Quantum}\ }\textbf {\bibinfo {volume} {4}},\ \bibinfo {pages}
  {010331} (\bibinfo {year} {2023})}\BibitemShut {NoStop}%
\bibitem [{\citenamefont {Lavasani}\ \emph {et~al.}(2022)\citenamefont
  {Lavasani}, \citenamefont {Luo},\ and\ \citenamefont {Vijay}}]{ali2022}%
  \BibitemOpen
  \bibfield  {author} {\bibinfo {author} {\bibfnamefont {A.}~\bibnamefont
  {Lavasani}}, \bibinfo {author} {\bibfnamefont {Z.-X.}\ \bibnamefont {Luo}},\
  and\ \bibinfo {author} {\bibfnamefont {S.}~\bibnamefont {Vijay}},\ }\bibfield
   {title} {\bibinfo {title} {Monitored quantum dynamics and the kitaev spin
  liquid}\ }\href {https://doi.org/10.48550/arxiv.2207.02877}
  {10.48550/arxiv.2207.02877} (\bibinfo {year} {2022})\BibitemShut {NoStop}%
\bibitem [{\citenamefont {Sriram}\ \emph {et~al.}(2022)\citenamefont {Sriram},
  \citenamefont {Rakovszky}, \citenamefont {Khemani},\ and\ \citenamefont
  {Ippoliti}}]{https://doi.org/10.48550/arxiv.2207.07096}%
  \BibitemOpen
  \bibfield  {author} {\bibinfo {author} {\bibfnamefont {A.}~\bibnamefont
  {Sriram}}, \bibinfo {author} {\bibfnamefont {T.}~\bibnamefont {Rakovszky}},
  \bibinfo {author} {\bibfnamefont {V.}~\bibnamefont {Khemani}},\ and\ \bibinfo
  {author} {\bibfnamefont {M.}~\bibnamefont {Ippoliti}},\ }\bibfield  {title}
  {\bibinfo {title} {Topology, criticality, and dynamically generated qubits in
  a stochastic measurement-only kitaev model}\ }\href
  {https://doi.org/10.48550/arxiv.2207.07096} {10.48550/arxiv.2207.07096}
  (\bibinfo {year} {2022})\BibitemShut {NoStop}%
\bibitem [{\citenamefont {Dua}\ \emph {et~al.}()\citenamefont {Dua},
  \citenamefont {Ellison}, \citenamefont {Sullivan},\ and\ \citenamefont
  {Tantivasadakarn}}]{ZNpaper}%
  \BibitemOpen
  \bibfield  {author} {\bibinfo {author} {\bibfnamefont {A.}~\bibnamefont
  {Dua}}, \bibinfo {author} {\bibfnamefont {T.~D.}\ \bibnamefont {Ellison}},
  \bibinfo {author} {\bibfnamefont {J.}~\bibnamefont {Sullivan}},\ and\
  \bibinfo {author} {\bibfnamefont {N.}~\bibnamefont {Tantivasadakarn}},\
  }\href@noop {} {\ }\bibinfo {note} {(to appear), (2022)}\BibitemShut
  {NoStop}%
\bibitem [{\citenamefont {Brown}\ and\ \citenamefont
  {Williamson}(2020)}]{Brown_2020}%
  \BibitemOpen
  \bibfield  {author} {\bibinfo {author} {\bibfnamefont {B.~J.}\ \bibnamefont
  {Brown}}\ and\ \bibinfo {author} {\bibfnamefont {D.~J.}\ \bibnamefont
  {Williamson}},\ }\bibfield  {title} {\bibinfo {title} {Parallelized quantum
  error correction with fracton topological codes},\ }\href
  {https://doi.org/10.1103/PhysRevResearch.2.013303} {\bibfield  {journal}
  {\bibinfo  {journal} {Phys. Rev. Res.}\ }\textbf {\bibinfo {volume} {2}},\
  \bibinfo {pages} {013303} (\bibinfo {year} {2020})}\BibitemShut {NoStop}%
\bibitem [{\citenamefont {Zhang}\ \emph {et~al.}(2022)\citenamefont {Zhang},
  \citenamefont {Aasen},\ and\ \citenamefont {Vijay}}]{zhang_2022}%
  \BibitemOpen
  \bibfield  {author} {\bibinfo {author} {\bibfnamefont {Z.}~\bibnamefont
  {Zhang}}, \bibinfo {author} {\bibfnamefont {D.}~\bibnamefont {Aasen}},\ and\
  \bibinfo {author} {\bibfnamefont {S.}~\bibnamefont {Vijay}},\ }\href
  {https://doi.org/10.48550/ARXIV.2211.05784} {\bibinfo {title} {The x-cube
  floquet code}} (\bibinfo {year} {2022})\BibitemShut {NoStop}%
\bibitem [{\citenamefont {Bonilla~Ataides}\ \emph {et~al.}(2021)\citenamefont
  {Bonilla~Ataides}, \citenamefont {Tuckett}, \citenamefont {Bartlett},
  \citenamefont {Flammia},\ and\ \citenamefont {Brown}}]{bonilla2021xzzx}%
  \BibitemOpen
  \bibfield  {author} {\bibinfo {author} {\bibfnamefont {J.~P.}\ \bibnamefont
  {Bonilla~Ataides}}, \bibinfo {author} {\bibfnamefont {D.~K.}\ \bibnamefont
  {Tuckett}}, \bibinfo {author} {\bibfnamefont {S.~D.}\ \bibnamefont
  {Bartlett}}, \bibinfo {author} {\bibfnamefont {S.~T.}\ \bibnamefont
  {Flammia}},\ and\ \bibinfo {author} {\bibfnamefont {B.~J.}\ \bibnamefont
  {Brown}},\ }\bibfield  {title} {\bibinfo {title} {The xzzx surface code},\
  }\href {https://www.nature.com/articles/s41467-021-22274-1} {\bibfield
  {journal} {\bibinfo  {journal} {Nature communications}\ }\textbf {\bibinfo
  {volume} {12}},\ \bibinfo {pages} {1} (\bibinfo {year} {2021})}\BibitemShut
  {NoStop}%
\bibitem [{\citenamefont {Tuckett}\ \emph {et~al.}(2019)\citenamefont
  {Tuckett}, \citenamefont {Darmawan}, \citenamefont {Chubb}, \citenamefont
  {Bravyi}, \citenamefont {Bartlett},\ and\ \citenamefont
  {Flammia}}]{PhysRevX.9.041031}%
  \BibitemOpen
  \bibfield  {author} {\bibinfo {author} {\bibfnamefont {D.~K.}\ \bibnamefont
  {Tuckett}}, \bibinfo {author} {\bibfnamefont {A.~S.}\ \bibnamefont
  {Darmawan}}, \bibinfo {author} {\bibfnamefont {C.~T.}\ \bibnamefont {Chubb}},
  \bibinfo {author} {\bibfnamefont {S.}~\bibnamefont {Bravyi}}, \bibinfo
  {author} {\bibfnamefont {S.~D.}\ \bibnamefont {Bartlett}},\ and\ \bibinfo
  {author} {\bibfnamefont {S.~T.}\ \bibnamefont {Flammia}},\ }\bibfield
  {title} {\bibinfo {title} {Tailoring surface codes for highly biased noise},\
  }\href {https://doi.org/10.1103/PhysRevX.9.041031} {\bibfield  {journal}
  {\bibinfo  {journal} {Phys. Rev. X}\ }\textbf {\bibinfo {volume} {9}},\
  \bibinfo {pages} {041031} (\bibinfo {year} {2019})}\BibitemShut {NoStop}%
\bibitem [{\citenamefont {Miguel}\ \emph {et~al.}(2023)\citenamefont {Miguel},
  \citenamefont {Williamson},\ and\ \citenamefont
  {Brown}}]{https://doi.org/10.48550/arxiv.2203.16534}%
  \BibitemOpen
  \bibfield  {author} {\bibinfo {author} {\bibfnamefont {J.~F.~S.}\
  \bibnamefont {Miguel}}, \bibinfo {author} {\bibfnamefont {D.~J.}\
  \bibnamefont {Williamson}},\ and\ \bibinfo {author} {\bibfnamefont {B.~J.}\
  \bibnamefont {Brown}},\ }\bibfield  {title} {\bibinfo {title} {A cellular
  automaton decoder for a noise-bias tailored color code},\ }\href
  {https://doi.org/10.22331/q-2023-03-09-940} {\bibfield  {journal} {\bibinfo
  {journal} {{Quantum}}\ }\textbf {\bibinfo {volume} {7}},\ \bibinfo {pages}
  {940} (\bibinfo {year} {2023})}\BibitemShut {NoStop}%
\end{thebibliography}%

\newpage

\clearpage
\pagebreak
\onecolumngrid

\appendix 
\setcounter{equation}{0}
\setcounter{figure}{0}
\setcounter{table}{0}
\makeatletter
\renewcommand{\theequation}{A\arabic{equation}}
\renewcommand{\thefigure}{A\arabic{figure}}

\section{Unitary circuit with measurements framework}

One difference between the honeycomb code and the CSS honeycomb code is that the ISG of the former always stays a subgroup of a fixed subsystem code and the latter one is not derivable from a subsystem code. Another way of comparing these two codes might be by noticing that the layout of the measurements is the same in both codes and thus, there must exist a depth-one unitary circuit relating the codes at a given time instance. This does not mean that there is an equivalence between the codes, because the unitary circuit is different at each time. 
In particular,  measuring $XX$ on a given link is equivalent to applying the unitary $HH$ on that link, then measuring $ZZ$ and then applying $HH$ again.  Similarly, measuring $YY$ on a given link is equivalent to applying the unitary $(SH)\otimes (SH)$, measuring $ZZ$ and then applying $(HS^{\dagger})\otimes (HS^{\dagger})$.

Therefore, we can reduce a Floquet code  to measurement of $ZZ$-checks of the color of the given round interspersed with a depth-1 unitary circuit.  For the CSS honeycomb code, one can verify that this depth-1 unitary circuit at each iteration is
\begin{equation}
    U_{\text{CSS}}= \bigotimes_{\bm x}H_{\bm x}
\end{equation}
where $\bm x$ indexes the coordinates of qubits on the lattice.  For the honeycomb code  the depth-1 unitaries change with period 3. We define $\bm a_1 = (0,3/2)$ and $\bm a_2 = (3\sqrt{3}/4,3/4)$ and the origin of the lattice to coincide with the bottom of one of the vertical red checks, and the correspondence between the edge orientation and the check flavor is
$$\begin{tikzpicture}
\draw (0,0)--(0,0.5) node[right] {$z$};
\draw (0,0)--(0.433,-0.25) node[right] {$y$};
\draw (0,0)--(-0.433,-0.25) node[left] {$x$};
\end{tikzpicture}
$$
At step when the red-colored checks are measured, it is:
\begin{equation}
    \begin{split}
    U_{\text{HH,r}}&= \bigotimes_{ \bm x}U( \bm x)
    \\
 U(\bm x) &= \left\{\begin{matrix*}[l]
SH, &&  \bm x = \bm x_{nm},\\ 
H, && \bm x = \bm x_{nm}+ \lp 0,\frac{1}{2}\rp,\\ 
H, && \bm x = \bm x_{nm}+  \lp \frac{\sqrt 3}{2},0 \rp,\\ 
HS^\dagger, &&  \bm x = \bm x_{nm}+  \lp \frac{\sqrt 3}{2},\frac{1}{2}\rp,\\ 
S, && \bm x = \bm x_{nm}+  \lp \frac{\sqrt 3}{4}, \frac{1}{4}\rp,\\ 
HSH, && \bm x = \bm x_{nm}+  \lp \frac{\sqrt 3}{4}, \frac{3}{4}\rp.
\end{matrix*}\right.
    \end{split}
\end{equation}
where $\bm x_{nm}= n \bm a_1 + m \bm a_2$. At rounds when green and blue checks are measured, the unitary layer is 
\begin{equation}
    \begin{split}
        U_{\text{HH,g}}&= \bigotimes_{ \bm x}U \lp  \bm x - \lp \frac{\sqrt 3}{2},0 \rp \rp,\\
        U_{\text{HH,b}}&= \bigotimes_{ \bm x}U\lp \bm x - \lp \frac{\sqrt 3}{4}, \frac{1}{4}\rp \rp.
    \end{split}
\end{equation}
Thus, we see that the structure is quite different for the two Floquet codes. The difference even in a depth-one single-qubit unitary layer can affect the threshold properties of a code under different (biased) noise models \cite{bonilla2021xzzx,PhysRevX.9.041031,https://doi.org/10.48550/arxiv.2203.16534}; therefore it would be interesting to benchmark the code(s) discussed here and compare their performance to that of the honeycomb code.  A classification scheme could exist for dynamic codes based on the algebra of the check operators, or based on the algebra of the depth-1 unitaries.

%This suggests a fundamental difference between the two codes, as the single depth unitaries have different periods in both codes, and the unitary for the CSS honeycomb code is site independent, while for the HH code it is site dependent.  We believe that more broadly, a classification scheme should exist for dynamic codes based on the algebra of the check operators, or equivalently the algebra of the depth-1 unitaries.

\section{Preparation protocol for Haah's code}\label{app:Haahprep}

\begin{figure}[ht]
\begin{center}
\begin{minipage}[t]{.5\linewidth}
\vspace{0pt}
\centering
\vspace{0pt}
		\includegraphics[width= 0.75\columnwidth]{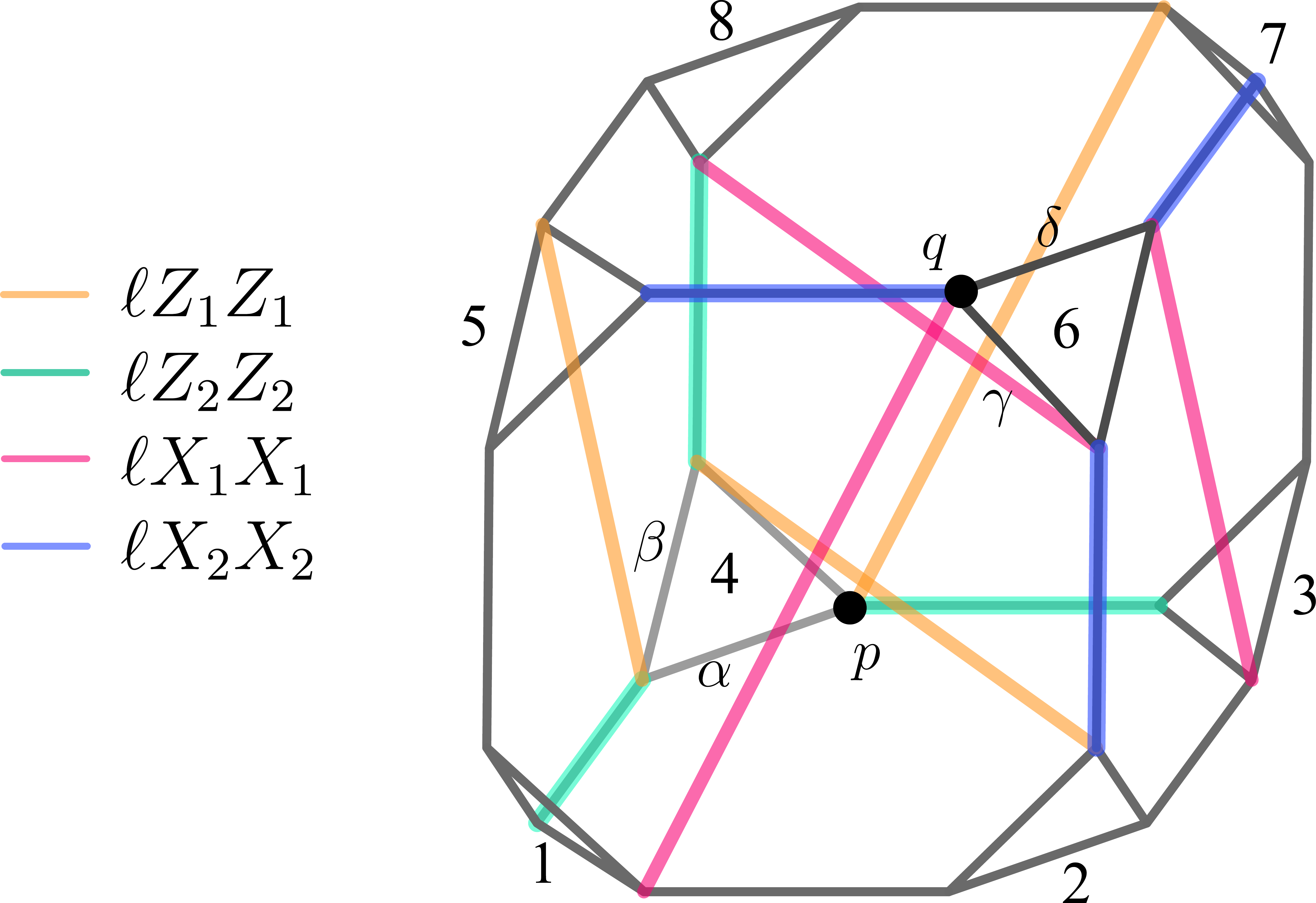}
\end{minipage}%
\begin{minipage}[t]{.5\linewidth}
\vspace{0pt}
\centering
   \begin{tabular}{p{0.3cm}|l}
$r$ &  $\mathcal{S}(r)$\\ \hline
0 &  $\ell XX, \ell ZZ$ on colored edges in figure\\
1 &  $(\alpha,\beta) X_{1}X_{1}$, $(\alpha,\beta) X_{2}X_{2}$,  $(\gamma,\delta) Z_{1}Z_{1}$, $(\gamma,\delta) Z_{2}Z_{2}$\\
 & 
$p X_1 X_2$, $q Z_1Z_2$, \text{Haah's code stabilizers}
\end{tabular}
\end{minipage}
\caption{Construction of Haah's code}
	 \label{fig8}
\end{center}
\end{figure}

This code prepares Haah's code in two layers via measuring two-qubit gauge checks.  However, we are not aware if it is possible to adapt this protocol in order to make it dynamical.  As in the checkerboard model preparation, we divide each site of the square lattice into 6 sites, forming an octahedron.  Each site on the decorated lattice hosts two spins now, of types `1' and `2'.  

We draw the configuration of links formed in Fig.~\ref{fig8}.  Since there are two links coming out of each site, each link corresponds to a two-spin interaction.  The preparation protocol is therefore shown in the table above. 
The notation $\ell$ corresponds to the links labelled in the figure, where we assume that orange and pink links act on  qubits of type `1', and blue and green links act on qubits of type `2'.  $\alpha, \beta, \gamma, \delta$ correspond to the labelled edges while $p$ and $q$ correspond to labelled vertices.  The subscripts $1$ and $2$ correspond to the flavors of the spins at each site. 

Note that per octahedron there are initially 12 qubits.  The checks of the round $r=2$, namely $\alpha X_{1,2}X_{1,2}$, $\beta X_{1,2}X_{1,2}$ and $\gamma Z_{1,2}Z_{1,2}$, $\delta Z_{1,2}Z_{1,2}$ force there to be 4 effective qubits left.
Next, two of these qubits are eliminated by measuring $p X_1X_2$ and $q Z_1Z_2$.  Additionally, after performing the checks of this round, the edge measurements of round $r=1$ form Haah's code stabilizers.

\end{document}